\documentclass[hyper]{JHEP3}
\pdfoutput=1
\usepackage{amsmath,amssymb,amsfonts,amsxtra,mathrsfs,makeidx,graphics,graphicx,amsthm,bbm}
\usepackage{epsfig}
\usepackage[all]{xy}

\newcommand{\SC}{{\mathsf C}}

\newcommand{\CA}{\mathcal{A}}

\newcommand{\CS}{\mathcal{S}}

\newcommand{\CL}{\mathcal{L}}

\newcommand{\CN}{\mathcal{N}}
\newcommand{\CO}{\mathcal{O}}

\newcommand{\CR}{\mathcal{R}}

\newcommand{\CD}{\mathcal{D}}

\renewcommand{\Im}{{\rm Im}}
\renewcommand{\Re}{{\rm Re}}
\newcommand{\Tr}{\mbox{Tr}}

\newcommand{\IR}{\mathbb{R}}

\newcommand{\IZ}{\mathbb{Z}}

\newcommand{\SU}{\mathrm{SU}}
\newcommand{\SO}{\mathrm{SO}}
\newcommand{\USp}{\mathrm{USp}}
\newcommand{\SL}{\mathrm{SL}}
\newcommand{\U}{\mathrm{U}}
\newcommand{\AdS}{\mathrm{AdS}}
\newcommand{\CFT}{\mathrm{CFT}}
\newcommand{\SCFT}{\mathrm{SCFT}}

\def\Z{\mathbb{Z}}

\newcommand{\Dslash}{\; \ensuremath \raisebox{0.03cm}{\slash}\hspace{-0.28cm} D}
\newcommand{\CDslash}{\; \ensuremath \raisebox{0.03cm}{\slash}\hspace{-0.28cm} {\cal D}}

\newcommand{\half}{\frac{1}{2}}
\newcommand{\ndt}{\noindent}

\newcommand{\pa}{\partial}

\def\Lt{{\wt L}}
\def\e{\epsilon}

\def\i{\mathrm{i}}

\def\p{\partial}

\def\bea{\begin{eqnarray}}
\def\eea{\end{eqnarray}}
\def\be{\begin{equation}}
\def\ee{\end{equation}}
\def\ba{\begin{align}}
\def\ea{\end{align}}
\def\bse{\begin{subequations}}
\def\ese{\end{subequations}}

\newcommand{\bem}{\begin{pmatrix}}
\newcommand{\eem}{\end{pmatrix}}

\def\zb {\bar{z}}

\def\wb {\bar{w}}

\def\={\;  = \;}
\def\+{\, + \,}

\def\wt{\widetilde}

\def\bar{\overline}

\def\rt2{\sqrt{2}}

\newcommand{\I}{\mathrm{i}}
\renewcommand{\Im}{\mbox{Im}}
\renewcommand{\Re}{\mbox{Re}}

\def\s{\sigma}
\def\g{\gamma}
\def\t{\tau}
\def\a{\alpha}
\def\b{\beta}
\def\d{\delta}

\def\G{\Gamma}

\newcommand{\sm}[4]{\bigl(\smallmatrix #1&#2\\ #3&#4\endsmallmatrix\bigr)}

\title{Which $\rm{\bf AdS_3}$ configurations contribute to \\ the $\rm{\bf SCFT_2}$ elliptic genus?}

\author{Sameer Murthy${}^{1}$ and Satoshi Nawata${}^{2}$\\
${}^1$Institute for Theoretical Physics, Utrecht University, \\
Leuvenlaan 4, 3584 CE Utrecht, The Netherlands\\
\\
${}^2$Perimeter Institute for Theoretical Physics, \\
Waterloo Ontario Canada, N2L 2Y5 \\

{\tt \email{S.V.Murthy@uu.nl, snawata@perimeterinstitute.ca}} }

\abstract{ 
According to the $\AdS/\CFT$ duality, the superconformal index of a 
superconformal field theory should have an $\AdS$ interpretation as a 
Euclidean functional integral with periodic boundary conditions on the fermions. 
Unlike the thermal case, the Euclidean continuation of the supersymmetric 
$\AdS$ black hole does not smoothly fill in these boundary conditions,
leading us to ask the title question. 
In the context of $\AdS_{3}/\CFT_{2}$, we show using supersymmetric localization that 
the gravitational functional integral for the elliptic genus localizes onto asymptotically 
$\AdS_{3}$ configurations that are annihilated by a certain supercharge, 
 in the relevant off-shell supergravity theory. 
For $(0,4)$ superconformal field theories, we find such a localizing configuration in the 
5d $\CN=2$ off-shell supergravity theory that is asymptotically $\AdS_{3} \times S^{2}$. 
This configuration interpolates smoothly between 
the supersymmetric BTZ black hole 
in the interior and a constant gauge field configuration at the boundary, thus 
smoothly filling in the $(++)$ boundary conditions 
for spinors on the boundary torus. It has an action equal to the 
supersymmetric BTZ black hole,  
holomorphic in the complex structure $\t$ of the boundary torus. 
Our results have interesting implications for the 
black hole Farey tail in $\AdS_{3}$, as well as for higher dimensional $\AdS$ theories. 
}


\begin{document}

\section{Introduction and summary}

One of the most interesting applications of the $\AdS/\CFT$ correspondence is to use 
calculations in the boundary field theory to gain insights about the dual quantum theory of 
gravity. Faced with the strong/weak coupling nature of the correspondence, a good strategy 
is to consider quantities that are protected by supersymmetry, and therefore 
do not change under a continuous change of the coupling constant. 
One computes such a quantity in a weakly coupled field theory regime and then attempts to 
interpret it in terms of the gravitational variables which are more natural in the opposite regime 
of strongly coupled field theory. 
Since we have quite a few exact finite charge results for the weakly coupled field theory 
computations in various dimensions, 
a successful application of this strategy can give detailed information about the 
quantum gravitational theory. 
Recently, this has been applied to the case of the $\AdS_{2}/\CFT_{1}$ case with 
a good degree of success \cite{Dabholkar:2010uh, Dabholkar:2011ec}, and it would 
be very interesting to extend this to the higher dimensional $\AdS_{d+1}/\CFT_{d}$ cases. 

One such quantity of interest, which will be the focus of this paper, is the supersymmetric index.
At an inverse temperature $\b$, the thermal partition function has the form $\Tr \, e^{-\b H}$,
where $H$ is the Hamiltonian, and the trace is over the Hilbert space of the theory. 
The supersymmetric index $\Tr \, (-1)^{F} e^{-\b H}$ is the partition function weighted by the 
fermion number. Depending on the details of the theory, it can also be weighted by chemical 
potentials which couple to conserved charges. 

The thermal partition function has an interesting interpretation 
in terms of the gravitational variables 
-- at low temperatures, the spectrum can be understood as a gas of 
gravitons, while at high temperatures, the dominant contribution to the partition function is  
from a black hole in $\AdS$ space \cite{Witten:1998zw}. 
It is natural to ask if a similar interpretation exists for the 
supersymmetric index, where one could hope to make exact  statements, 
this is the question posed in the title. 
In particular, does the $\SCFT_{d}$ superconformal index have a gravitational interpretation 
as containing a supersymmetric black hole\footnote{By a supersymmetric 
black hole, we mean the supersymmetric limit of the corresponding non-extremal configuration, 
this allows us to cleanly compute its contribution to the functional integral \cite{Sen:2007qy}.}
in $\AdS_{d+1}$ for some range of chemical potentials 
(similar to the high temperature thermal partition function)?

In order to make a more precise investigation, we shall use the Euclidean functional integral formalism. 
We fix the boundary conditions to the classical $\AdS_{d+1}$ configuration 
at the boundary, and we would like to integrate over all fluctuations in the 
interior\footnote{Since the theory includes gravity, there is, of course, the usual problem with 
the ultraviolet behavior of the theory -- the fluctuations grow too fast at high energies and the 
naive functional integral is not convergent. Nevertheless, by assuming that there is some theory, 
like string theory, which cuts off the high energy fluctuations in an appropriate way, we can 
try to make sense of the semiclassical limit of the functional integral.}. 
Here we run into a puzzle: the Euclidean $\AdS$ black hole has a contractible time circle 
in the interior, and hence the only allowed smooth boundary condition for the fermions 
is the anti-periodic ($-$) one. This is consistent with the interpretation that the black hole 
dominates the thermal functional integral which also has $(-)$ boundary conditions.  
On the other hand, the supersymmetric index is computed as a functional integral 
with periodic $(+)$ boundary conditions for the fermions. 
We are thus led to conclude that:
\begin{quote}
\emph{the supersymmetric Euclidean AdS black hole  cannot contribute 
to the AdS gravitational functional integral dual to the supersymmetric index. }
\end{quote}
One may try to remedy this by turning on flat gauge fields around the time circle, 
but this does not work, since one cannot smoothly turn on a flat gauge field 
around a contractible circle. 
This conclusion seems to be at odds with a general belief based on some data (which 
we review below), that the $\SCFT_{2}$ index, interpreted as a bulk $\AdS_{3}$ object, 
generically has a black hole like behavior. In this paper, we shall sharpen and resolve this puzzle.

The resolution of this puzzle is part of our more general program to exactly evaluate the supersymmetric 
functional integral of a gravitational theory on $\AdS_{3}$. 
We will consider the elliptic genus of a $\SCFT_{2}$ \cite{Schellekens:1986yi, 
Schellekens:1986xh, Pilch:1986en, Witten:1986bf, Witten:1987cg, Alvarez:1987wg,Alvarez:1987de}, 
which needs at least $(0,2)$ supersymmetry\footnote{$\CN=1$ supergravity on $\AdS_{3}$ 
has been discussed in \cite{Maloney:2007ud}.}. 
Euclidean $\AdS_{3}$ is topologically a solid torus, with one of the circles being contractible
in the interior. In the functional integral formalism, the elliptic genus computation is done with $(++)$ 
boundary conditions for the spinors around the two circles at the boundary.

The idea of an exact evaluation of the $\AdS_{3}$ functional integral dual to the elliptic genus 
of $\SCFT_{2}$ was proposed in the seminal paper \cite{Dijkgraaf:2000fq}, and 
progress on the idea was made in \cite{Kraus:2006nb} (see \cite{Kraus:2006wn} for a nice review). 
The bulk configurations in \cite{Dijkgraaf:2000fq} had flat gauge fields turned on around this 
contractible circle, 
and the induced delta function singularity at the center of $\AdS_{3}$ was interpreted
as coming from point-like sources. Considering that the interpretation of the 
thermal partition function does not need 
any such sources, this situation is somewhat unsatisfactory, and it is interesting to 
ask if there are any gravitational configurations which smoothly fill in the $(++)$ boundary 
conditions.

A related problem which was left open in \cite{Dijkgraaf:2000fq} and its follow-ups is the following: 
the elliptic genus of a $(0,2)$ $\SCFT_{2}$ 
is known to be holomorphic in the complex structure $\t$ of the torus on which it is defined, 
since the fermion number $(-1)^{F}$ pairs the massive right-moving modes.
What is the mechanism in the dual $\AdS_{3}$ problem by which the bulk functional integral becomes 
holomorphic in the complex structure $\tau$ of the boundary torus? 
This question can be reformulated in the following manner\footnote{This was 
emphasized to us by Ashoke Sen, for which we thank him.} -- what is saddle point of 
the $\AdS_{3}$ Euclidean functional integral with the boundary torus having 
a generic complex structure $\tau$? Such a torus has $\bar \tau = \tau^{*}$, and 
so the answer cannot be the  Euclidean continuation of the extremal BTZ black hole \cite{Banados:1992wn}
which has $\tau = M+iJ$ finite and $\bar \tau = M-iJ=0$. 

In this paper, we shall resolve both these puzzles. 
In the rest of this introduction, we shall expand on these related puzzles, 
sharpen them by including a consideration of gauge fields, and explain our strategy
to solve the problem. 

\subsection{A more detailed consideration of the puzzle}

\ndt \emph{Entropy considerations in $\SCFT_2$}

We first review some data that seem to point to a black hole interpretation of the 
$\AdS_{3}$ index. 
The elliptic genus of the $\SCFT_{2}$ is a generating function for the number of 
(bosonic $-$ fermionic) states in the Ramond sector annihilated by the right-moving supercharges. 
The right-movers are therefore  in the Ramond sector ground state ($\wt L_{0}~\!\!-~\!\!\wt c/24\!\!~=~\!\!0$), 
and there are non-trivial excitations in the left-moving sector with energy $L_{0}$.  
Their degeneracy can be estimated by the Cardy formula to have an exponential growth 
of states $\Omega~\approx~\exp(2\pi \sqrt{c L_{0}/6})$, 
where $c$ is the central charge of the $\SCFT_2$. 

In the dual $\AdS_{3}$, the supersymmetric BTZ black hole preserves 
the right-moving supersymmetry of the original $\AdS_{3}$ theory, and has 
non-negative mass $M$ ($=$ spin $|J|$), so that $L_{0} -\frac{c}{24}= M, \; \wt L_{0}-\frac{\wt c}{24}=0$. 
Further, it has an entropy given by $S = k_{B} \log \Omega$. 
It is thus natural to think that the black hole represents the
collection of supersymmetric states in the boundary theory counted by the elliptic genus, 
{\it i.e.} states in the R sector vacuum\footnote{For $(2,2)$ theories,
the RR vacuum would be the ``zero mass'' ($L_{0}-c/24=0$) BTZ black hole.} 
\cite{Strominger:1998yg}. 
As mentioned above, a more refined version of this statement has been envisioned in the {\it black hole Farey tail} 
\cite{Dijkgraaf:2000fq}, wherein the exact elliptic genus would be accounted for by the BTZ 
black hole and fluctuations around it, summed over the $\SL(2,\IZ)$ family of black holes 
\cite{Maldacena:1998bw} in the Euclidean theory.

By contrast, in $\AdS_{4}$ and $\AdS_{5}$, the most general supersymmetric index 
defined using the superalgebra of the theory grows too slowly  \cite{Kinney:2005ej,Bhattacharya:2008zy, Kim:2009wb} to accommodate the 
density of states of the corresponding supersymmetric $\AdS$ black holes 
\cite{Gutowski:2004ez, Gutowski:2004yv, Chong:2005hr}. Our conclusion that black holes 
do not contribute to the index seems to be consistent\footnote{In general, entropy considerations like the above 
are not conclusive since one is comparing the microscopic index which counts 
the difference in the number of bosons and fermions, with the macroscopic black hole 
entropy which should represent a count of the absolute number of states. However, the 
entropy is never less than the index for a given set of charges, and 
our puzzle for $\AdS_{3}$ remains.} with the entropy data of higher 
dimensions, but not three dimensions. 
The case of the lower dimensional $\AdS_{2}$ is special since, as we review below, 
the gravitational entropy and index are in fact equal \cite{Sen:2009vz, Dabholkar:2010rm}. 
Indeed, there has indeed been impressive 
agreement of the $\AdS_{2}$ partition function with the microscopic index, well beyond the 
infinite charge expansion (\cite{LopesCardoso:1998wt, LopesCardoso:2004xf, Castro:2008ys}, 
see \cite{Sen:2007qy} for a review, and very recently, \cite{Dabholkar:2010uh, Banerjee:2011jp, 
Sen:2011ba, Sen:2011cj, Dabholkar:2011ec}).

\vspace{0.2cm}

\ndt \emph{Gauge fields in the functional integral}

One may try to save the black hole interpretation in $\AdS_{3}$ 
by turning on a gauge field under which the fermions of the theory are charged, in 
the background of the black hole. A gauge field configuration which has a non-zero integral 
around the time circle at infinity will indeed change the effective periodicity of the fermions which 
couple to it, but such a configuration will necessarily enclose flux due to the contractibility of the circle. 
In the Lorentzian theory, one can turn on a smooth flat gauge field configuration 
\cite{Maldacena:2000dr},  
but its Euclidean continuation will lead to a singularity at the origin \cite{deBoer:2008fk}. 
If we demand that the gauge field configuration be smooth, it will have a non-zero field strength,
and it will backreact to destroy the black hole geometry. 

Since we would eventually like to go beyond the saddle point approximation and evaluate 
exact functional integrals, it is important to fix the ensemble in which we work. 
This is specified by the choice of boundary conditions for the $\AdS$ functional integral. 
Near the boundary, where the fields can be taken to be approximately free, 
the Maxwell field has two independent solutions -- the 
gauge potential and the electric field. 

In $\AdS_{2}$ the electric field mode which carries the charge dominates the gauge 
potential mode near the boundary. This implies that the gauge potential mode is integrated over, 
and hence the value of $\oint A$ around the time circle at infinity can be changed by a shift of 
the integration variable in the functional integral\footnote{We thank Ashoke Sen for explaining 
this point to us in detail. The argument can also be understood quite simply in the Hamiltonian 
formalism \cite{Sen:2009vz, Dabholkar:2010rm}. One naturally has the \emph{microcanonical ensemble} 
where all the charges, including the one associated to the fermion number current $F$ are held fixed. 
This implies that $(-1)^{F}$ has a constant value over all the states of the ensemble and so $\Tr \, (-1)^{F}$ is 
proportional to $\Tr \, 1$. Thus, upto a constant, the supersymmetric index equals the entropy of the system.}.
In practice therefore, one can work with the smooth $\AdS_{2}$ solution as a saddle point. 
In contrast, in $\AdS_{d > 3}$, the dominant mode is the gauge potential, while the electric field mode
is sub-dominant. Fixing the gauge potential brings us to 
the \emph{canonical ensemble} where the charge is allowed to fluctuate. 
The effective periodicity of the fermions cannot be changed by a normalizable deformation. 

The case of $\AdS_{3}$ is subtle since the long distance theory is described by 
Chern-Simons theory with a first order equation of motion. If one fixes the radial 
component of the gauge field to be zero by a gauge choice, 
the other two components are canonical conjugates, and 
one should not fix both their values at the boundary. Instead, the prescription 
\cite{Witten:1988hf, Elitzur:1989nr} is to fix one of these legs of the gauge field 
and allow the other to fluctuate in the interior. For the elliptic genus computation, 
one must fix the mode of the gauge field which couples to the holomorphic chemical
potential \cite{Dijkgraaf:2000fq}.  

If we now demand that gauge fields at infinity have the standard reality condition imposed on them, 
namely that the conjugate chemical potentials obey 
$\bar z= z^{*}$, this fixes the values of both the legs of the gauge field at infinity.  
This can be restated in the boundary theory as follows. In order to go from the NS 
to the R sector in the boundary $\SCFT_2$ (this operation is often referred to as spectral flow), 
one has to change the chemical potential $z$, which is clearly part of the fixed data of the 
boundary theory. The bottom line is that in $\AdS_{3}$, the gauge fields, and therefore 
the periodicity of the fermions, are fixed at infinity and cannot be changed by a fluctuation 
in the quantum theory. Our results can also be thought of as finding a smooth mechanism 
for spectral flow in the bulk.

\subsection{Strategy and results}

\ndt \emph{Supersymmetric localization}

Our strategy to answer the questions we posed above is based on the technique of 
supersymmetric localization \cite{Witten:1988ze, Witten:1991zz, Witten:1991mk, Schwarz:1995dg}. 
We shall argue that the $\AdS_{3}$ functional integral can be 
localized to a set of configurations which are annihilated by a certain supercharge of the theory. 
These configurations will, in general, not solve the equations of motion of the theory, 
and we shall need to use an off-shell version of supergravity including the auxiliary fields 
in which the supersymmetry variations close without using the equations of motion of the 
theory\footnote{On-shell supergravities on $\AdS_{3}$ have been explored earlier 
in \cite{Achucarro:1987vz} and \cite{David:1999nr}.}.

We investigate the case of a $\SCFT_{2}$ with $(0,4)$ supersymmetry, which has 
an $\SU(2)$ $R$-symmetry. 
Accordingly, we shall place ourselves in the context of 5d supergravity with 
$\AdS_{3} \times S^{2}$ boundary conditions with $(++)$ fermion boundary conditions 
on the boundary torus. 
In the purely gravitational sector, we find a smooth solution to the localization equations which 
has the following features. Near the boundary, the configuration is $\AdS_{3}$ with a constant 
gauge field potential which couples to the $\wt J_{3}$ component of the $\SU(2)$ $R$-symmetry. 
Near the origin, the configuration is a supersymmetric BTZ black hole whose geometry is determined 
by the equations of motion of the theory.
The localizing solution has a non-trivial fibration of the $S^{2}$ over the $\AdS_{3}$, and 
smoothly interpolates between the supersymmetric black hole near the origin and 
the torus at infinity, but it does not obey the equations of motion in the intermediate region. 
The action of this configuration is exactly equal to that of the supersymmetric black hole. 
We consider this to be the correct {\it universal} starting point for the 
index calculation in $\AdS_{3}$ supergravity, analogous to how the black hole saddle point
is a good starting point for the thermal partition function. 

Our configurations are essentially a lift of the off-shell BPS configurations in 
$\AdS_{3}$ found in \cite{Izquierdo:1994jz}, to five dimensions. 
These configurations did not have a physical meaning in the three dimensional theory of \cite{Izquierdo:1994jz}, 
since one needed the equation of motion to close the superalgebra \cite{Achucarro:1987vz}. 
Here, they find a genuine life as supersymmetric configurations of the 5d 
off-shell $\CN=2$ supergravity with non-trivial auxiliary fields. 

\vspace{0.2cm}

\ndt \emph{Plan of the paper}

In \S\ref{ads3partfn}, we review facts about gravitational theories on $\AdS_{3}$, and the interpretation 
of the gravitational partition function as that of a boundary $\CFT_{2}$. 
We also sharpen the problem with the spectral flow mechanism in the bulk. 
In \S\ref{localization},  we review the superalgebra on $\AdS_{3}$, and the supersymmetric index 
built using this algebra. We then briefly review the technique of supersymmetric localization, and apply it 
to a theory on $\AdS_{3}$ space. 
In \S\ref{5dreview}, we briefly review the $\CN=2$ off-shell supergravity in five dimensions coupled 
to an arbitrary number of vector multiplets, and present the maximally supersymmetric
$\AdS_{3} \times S^{2}$ solution of this theory. 
In \S\ref{localizingsolns}, we analyze the localizing equations and find the 
off-shell BPS localizing configurations which have the $(++)$
spinor boundary conditions. 
In \S\ref{holomorphy}, we analyze these solutions to get a physical picture. 
In \S\ref{discussion}, we end with a discussion of the implications of our solutions and  
interesting future directions to take. 
In four appendices, we discuss various technical details of our calculations.

\section{Partition functions on $\AdS_{3}$ \label{ads3partfn}}

In this section, we review some facts about gravitational theories on $\AdS_{3}$. 
In \S\ref{ads3basics}, we shall present the effective action and the solutions corresponding 
to pure $\AdS_{3}$ and the BTZ black hole.
We shall then discuss the asymptotic structure and symmetries of the space. 
In \S\ref{thermalBTZ}, we shall discuss the partition function of the theory, interpreted 
as a gravitational functional integral on $\AdS_{3}$, including a brief discussion of 
the relation between the thermal $\AdS_{3}$ and the BTZ black hole. 
In \S\ref{ellgen}, we shall discuss the puzzles that arise while considering the elliptic genus 
and spectral flow. 
In the first two subsections, we shall mostly follow the presentation of the review 
\cite{Kraus:2006wn}, and the related papers \cite{Kraus:2006nb,Hansen:2006wu}.

\subsection{Gravitational theories on $\AdS_{3}$ \label{ads3basics}}

We shall consider theories of the metric, a collection of gauge fields, scalar fields, and 
fermions. 
The three-dimensional bosonic effective action for the metric is  
\be \label{Sgrav}
{\cal S}_{\rm grav} = \frac{1}{16\pi G_3} \int_{\AdS_3} d^{3}x \sqrt{g} \, \big(R-\tfrac{2}{\ell^2} \big)
- \frac{1}{8\pi G_3} \int_{\p \AdS_3} d^{2}x \, \sqrt{h} \, \big(K-\tfrac{1}{\ell} \big) +
\cdots , 
\ee
We have explicitly written the two derivative action containing the Einstein-Hilbert term 
with cosmological constant and the Gibbons-Hawking boundary term. 
The ellipses indicate possible higher derivative terms, but we shall not consider them in this paper. 

The two derivative action has the $\AdS_{3}$ solution:
\be \label{ads3sol}
ds^2=- \left(1+ r^{2}/\ell^{2} \right)dt^2+\frac{dr^2}{1+r^{2}/\ell^{2}}+r^2d\phi^2 \, , 
\ee
a space of constant negative curvature which we have written in global coordinates. The radial 
coordinate $r$ runs from 0 to $\infty$, $r = \infty$ being the boundary of the space, and the 
angular coordinate $\phi$ is periodic with period $2 \pi$. 
This global $\AdS_{3}$ configuration plays the role of the vacuum of the theory. In the quantum theory, we should 
consider fluctuations around this space which preserve the asymptotic $\AdS_{3}$ boundary conditions. 
More precisely, expressing the metric in the Fefferman-Graham form \cite{Fefferman:1985}: 
\be \label{FeffGr}
g_{ij} = r^{2} g^{(0)}_{ij} + g^{(2)}_{ij} + {\cal O}\left(\tfrac1r\right) \, , 
\ee
we should keep $g^{(0)}_{ij}$ fixed, and allow the subleading terms to fluctuate. 
$g^{(0)}_{ij}$ is the conformal boundary metric, which shall be identified with the 
metric on the space on which the boundary $\CFT_2$ lives.

A two-parameter family of solutions of the effective action \eqref{Sgrav} which 
obey the boundary conditions \eqref{FeffGr} is given by the rotating 
BTZ black hole solutions \cite{Banados:1992wn}. These solutions are labeled by 
the radii $r_{\pm}$ of the inner and outer horizon
\be \label{BTZ}
ds^2=-\frac{\big(r^2-r_{+}^2\big)\big(r^2-r_{-}^2\big)}{r^2 \ell^{2}}dt^2
+\frac{\ell^{2} r^2}{\left(r^2-r_{+}\right)\left(r^2-r_{-}^{2}\right)}  dr^2 
+ r^2\left( d\phi+\frac{r_{+} r_{-}}{\ell r^2}dt \right)^2~.
\ee 
The mass and angular momentum of the black hole are related to the horizon radii as:
\be \label{MJBTZ}
M=\frac{r_+^2+r_-^2}{8G_{3} \ell^2} \, , \qquad J= \frac{r_+r_-}{4 G_{3} \ell} \, . 
\ee
The extremal BTZ black hole has $r_{+}=r_{-}$, or, equivalently, $M\ell=J$.

\vspace{0.2cm}

\ndt {\it Virasoro algebra}

As was discovered by Brown and Henneaux  \cite{Brown:1986nw}, gravitational theories 
on $\AdS_{3}$ have an infinite dimensional symmetry algebra, equivalent to two copies 
of the Virasoro algebra with central charges $(c, \wt c)$. This algebra can be identified 
with the left- and right-moving Virasoro algebra of the boundary $\CFT_{2}$.  For two derivative 
theories, the central charges are equal and can be expressed in terms of the $\AdS$ radius 
and the 3d Newton constant as $c = \wt c = {3\ell/2G_3}$.

The global charges $(L_{0,\pm1}, \wt L_{0,\pm1})$ are the generators of the finite dimensional 
algebra $\SO(2,2) = \SL(2,\IR)_{L} \times \SL(2,\IR)_{R}$ which are the isometries of the $\AdS_{3}$ 
geometry \eqref{ads3sol}. 
The mass and angular momentum in $\AdS_{3}$ are related to the Virasoso charges as:
\be\label{L0MJrel}
L_{0} - \frac{c}{24} = \half \big( M \ell + J \big) \, , \quad \wt L_{0} - \frac{\wt c}{24} = \half \big( M \ell - J \big) \, .
\ee
One can write these charge generators in terms of the geometric variables, as:
\be \label{BHVir}
L_{0} - \frac{c}{24}  = \frac{c}{48 \pi}\Big(\frac{d}{dt} - \frac{1}{\ell} \frac{d}{d\phi} \Big) +{\cal O}\big(\tfrac{1}{r^{4}} \big) \, , \qquad 
\wt L_{0} - \frac{\wt c}{24} = \frac{c}{48 \pi}\Big(\frac{d}{dt} + \frac{1}{\ell} \frac{d}{d\phi} \Big) + {\cal O}\big(\tfrac{1}{r^{4}} \big) \, .
\ee
One can also write explicit expressions for the full set of geometric Brown-Henneaux Virasoro generators 
\cite{Brown:1986nw}, but we will not need them here.

For an on-shell theory of gravity, the boundary stress tensor 
can be defined as the 
on-shell variation of the action with respect to the boundary metric. For a metric 
configuration of the form \eqref{FeffGr}, one has  \cite{Balasubramanian:1999re}:
\be\label{Tgrav}
T^{\rm grav}_{\a\b} = \frac{1}{8 \pi G \ell} \left(g^{(2)}_{\a\b} - (\Tr^{(0)} g^{(2)}) \, g^{(0)}_{\a\b}  \right) 
+ {\rm higher} \; {\rm derivative} \, .
\ee
In the following, we will need the Virasoro charges of the classical solutions \eqref{ads3sol} and 
\eqref{BTZ}, which we can compute using the above considerations. We have that  
the pure $\AdS_{3}$ metric has $L_{0} = \wt L_{0} = 0$, and for the 
extremal BTZ black hole with $M\ell =J$ we have $\wt L_{0} = \wt c/24$, $L_{0} = M\ell + c/24$.

\vspace{0.2cm}

\ndt \emph{Gauge fields} 

In this paper, we shall deal with left-moving $\U(1)$ gauge fields $A^{I}$, and a right-moving 
$\U(1)$ gauge field $\wt A$ (which is the third component of an $\SU(2)$ gauge field). 
A generic gauge field $A$ on $\AdS_{3}$ admits an expansion analogous to \eqref{FeffGr}
\be \label{GaugeFeffGr}
A = A^{(0)} + \tfrac{1}{r^{2}} A^{(2)} + {\cal O}\left(\tfrac1{r^{3}}\right) \, , 
\ee
we shall choose the gauge $A_{r}=0$. 
The leading long-distance action for the gauge fields is given by the Chern-Simons term 
\be\label{Sgauge}
{\cal S}_{\rm gauge} = \frac{i}{8\pi} \int_{\AdS_3} d^{3} x \, \big( k^{IJ} A_{I} \, dA_{J} - \wt k \wt A \, d \wt A \big)  
- \frac{1}{16\pi} \int_{\p \AdS_{3}} d^{2} x \, \sqrt{g} g^{\a \b} \big( k^{IJ} A_{I\a} \, A_{J\b} 
+ \wt k \, \wt A_{\a} \,  \wt A_{\b} \big)  \, .
\ee
The boundary term is obtained by demanding a consistent variational principle \cite{Kraus:2006nb}. 
The quantities $\wt k$ and $\wt c$ are related by supersymmetry as $\wt c = 6 \wt k$.

Since the boundary term in \eqref{Sgauge} depends on the metric, it contributes to the stress-energy tensor as: 
\bea \label{Tgauge}
T^{\rm gauge}_{\a \b}  =  \frac{k^{IJ}}{8 \pi} \left(A^{(0)}_{I\a} \, A^{(0)}_{J\b} -\frac{1}{2} A^{(0)\g}_{I}A^{(0)}_{J\g} \, 
g_{\a\b} \right)
+ \frac{\wt k}{8 \pi} \left(\wt A^{(0)}_{\a} \wt A^{(0)}_{\b} -\frac{1}{2} \wt A^{(0)\g} \wt A^{(0)}_{\g} g_{\a\b} \right) \, .
 \eea
Boundary currents which couple to the gauge fields are defined as 
the variation of the action with respect to the boundary value 
of the gauge fields. For the above action \eqref{Sgauge}, we get: 
\bea \label{current}
J^{I}_{\a}  =  \frac{i k}{4} \big(A^{(0)}_{I\a} - i \epsilon_{a}^{\b} A^{(0)}_{I\b} \big)  \, , \qquad 
\wt J_{\a}  =  \frac{i \wt k}{4} \big(\wt A^{(0)}_{\a} - i \epsilon_{a}^{\b} \wt A^{(0)}_{\b} \big)  \, . 
\eea

In the later sections of the paper, we shall consider 5d supergravity 
whose bosonic two derivative action is of the form
\be
\mathcal{S}^{5d}=  \frac{1}{4 \pi} \int\! d^5x \sqrt{g} \,  \left(\CR+\frac{1}{4} G_{IJ}F^I_{\mu\nu} F^{J\mu\nu} \right)  
  +\frac{i}{24 \pi} \int \! C_{IJK} \, W^I \wedge F^J \wedge F^K \ ,
  \label{5d action with CS}
\ee
where $C_{IJK}$ is a symmetric tensor. On compactification of this theory on $S^{2}$, 
one obtains a theory on $\AdS_{3}$. This theory has left-moving gauge fields $A^{I}$ 
coming from the 5d gauge fields $W^{I}$, and a right-moving 
gauge field $\wt A$ from the Kaluza-Klein reduction, which couples to the $R$-symmetry 
current of the dual boundary $\SCFT_2$. 
The Chern-Simon term for the gauge field $\wt A$ has a coefficient $\wt k$ which is determined by the 
tensor $C_{IJK}$ and the magnetic charges $p^{I}$ as $\wt k = \frac23 C_{IJK} p^{I} p^{J} p^{K}$.

\subsection{The partition function of the thermal $\AdS_{3}$ and the BTZ black hole \label{thermalBTZ}}

The $\AdS_{3}/\CFT_{2}$ dictionary states that the partition function of the 
$\AdS_{3}$ is equal to the partition function of the dual $\CFT_{2}$. The 
$\CFT_{2}$ is defined on the boundary of $\AdS_3$, which we take to be
a flat torus of modular parameter $\t$. 
We introduce the Euclidean time coordinate $t_{E} = - i t$,
and identify the boundary torus $T^2_{\rm bdry}$ coordinate as $w = \phi + it_{E}$. 
The metric on this torus $T^2_{\rm bdry}$ has a line element 
$ds^{2}= dw \, d\wb$, where the complex coordinate $w$ has a periodicity 
\be\label{wtorus}
w\sim w+2\pi \sim w + 2 \pi \tau \; .
\ee 
This boundary metric is identified with the leading value $g^{(0)}_{ij}$ of the bulk metric in \eqref{FeffGr}. 
In this manner, the complex structure $\t$ sets the boundary condition for the $\AdS_{3}$ metric.

The $\CFT_2$ partition function can be written in the Hamiltonian form as:
\be\label{CFTtrace}
 Z_{\rm CFT_2} = \Tr \left[ e^{2\pi i \tau ( L_0 - {c\over 24} ) -
2\pi i \bar\tau ( \wt L_0 -  {\wt c \over 24})} \right] \, .
\ee
If the theory contains fermions, we need to specify their periodicities around the circles 
of the torus. 
The periodicities around the space circle are imposed by restricting to the R $({\rm periodic}, +)$ 
or NS $({\rm antiperiodic}, -)$ sector. As written, \eqref{CFTtrace} implies anti-periodic boundary conditions 
around the  time circle. In order to impose a periodic boundary condition, we need 
to insert a $(-1)^{F}$ operator in the trace, where $F$ is the fermion number. 
By using the relation \eqref{L0MJrel}, 
we see that $\Im(\t)$ plays the role of inverse temperature, while $\Re(\t)$ is a chemical 
potential for the $\AdS_{3}$ angular momentum.

We would like to interpret $Z_{\rm CFT_2}$ as the partition function of the 
gravitational theory on $\AdS_{3}$. The conventional understanding of the Euclidean functional integral 
takes the form
\be \label{ZeS}
Z_{\rm \AdS_3} (\t, \bar \t) = \sum e^{-\CS} \, , 
\ee
where the summation runs over each saddle point of the action, and $\CS$ is the local 
effective action around the saddle point including the quantum fluctuations of massive modes.

The simplest saddle point is simply the thermal $\AdS_3$, which is obtained by identifying the 
imaginary time direction $t \sim t + i \b$ in the Lorentzian solution \eqref{ads3sol}. 
The Euclidean solution
is topologically equivalent to a solid torus, whose boundary is the torus $T^2_{\rm bdry}$ with
complex structure $\t$ as above. 
A second saddle point is the Euclidean continuation of the BTZ black hole \eqref{BTZ}. 
The Euclidean continuation is performed by  analytically continuing both the time coordinate
as above, and the parameter $r_- \to i r^{E}_-$, and then letting $t_{E}, r^{E}_-$ be real.
Regularity of the Euclidean section at $r_+$ requires identifying
\be
\label{perT}
(t,\phi) \sim \left( t+ \frac{\I}{\mathfrak{T}}, \phi+\frac{\I \mathfrak{J}}{\mathfrak{T}}\right)  \ ,\qquad
\mathfrak{T}=\frac{r_+^2-r_-^2}{2\pi \ell r_+}\ ,\quad \mathfrak{J} = \frac{r_-}{r_+} \, .
\ee
The Euclidean section is a solid torus filled with a three-dimensional hyperbolic metric. 
The A-cycle $(t(s), \phi(s))=(t_0 + \I s /\mathfrak{T},\phi_0+\I \mathfrak{J} s/\mathfrak{T})$ with $0\leq s<1$ is contractible in the
full geometry, and hence identified as the thermal circle, 
while the B-cycle $(t(s), \phi(s))=(t_0,\phi_0+ 2\pi s)$ is non-contractible.
The complex structure of the torus $T^2_{\rm bdry}$ generated by $\pa_{t_E}, \pa_\phi$
at fixed radial distance $r$ is parametrized by the modulus \cite{Murthy:2009dq}
\be\label{taurhop}
\tau_{+} = \frac{\I}{\ell}\left( r_{-} + r_+ \sqrt{\frac{r^2-r_-^2}{r^2-r_+^2}}\right)\, .
\ee
We define 
\be\label{taurhom}
\tau_{-} = \frac{\I}{\ell}\left( r_{-} - r_+ \sqrt{\frac{r^2-r_-^2}{r^2-r_+^2}}\right) \ ,
\ee
such that $\tau_{+}$ and $\tau_{-}$ are complex conjugate to each other when $r_-$ 
(and hence the angular momentum) is imaginary.
At large radius, the complex structure of the induced metric on the torus goes to a constant, 
\be\label{tauinf}
\tau^{\infty}_{\pm} =\frac{\I}{\ell}\ (r_{-} \pm r_+)\ .
\ee
In the Euclidean theory, one identifies $\tau^{\infty}_{\pm}$ with $(\tau, \bar \tau)$, which are 
complex conjugates of each other, after rotating $r_{-}$ to the imaginary axis. 

The above Euclidean BTZ solution is related to the thermal $\AdS_3$ by a modular transformation 
$\tau\to-1/\tau$ \cite{Maldacena:1998bw}. In fact, there is an $\SL(2,\Z)$ family of Euclidean solutions 
which can be obtained by performing modular transformations 
$\tau\to\gamma\cdot\tau= \frac{a\tau+b}{c\tau+d}$ on the complex structure $\tau$, where 
$\gamma\equiv \sm abcd  $ is an element of $\G\equiv \SL(2,\Z)$. 
These geometries are in one-to-one correspondence with the left coset $\G_\infty \setminus \G$ where the 
group $\G_\infty $  is the parabolic subgroup of translations and is generated by $\tau\to\tau+1$. 
All the known regular and black hole geometries arising as solutions of the Einstein equations with a negative 
cosmological constant in three dimensions can be obtained in this way. 
It follows that  the path integral in the leading order approximation can be written as the sum over all saddle points
\be
Z _{\rm grav}\sim \sum_{\g\in\G_\infty \setminus \G}  e^{-\mathcal{S}(\gamma\cdot \tau)}~.
\ee

We shall now present a quantitative discussion in a slightly more general setting, by
including a set of left-moving $\U(1)$ charges, one right-moving $\U(1)$ charge (which
will be the $R$-charge $\wt J^{3}_{0}$ in the $(0,4)$ setting), and their corresponding chemical potentials. 
The $\CFT_2$ partition function with left- and right-moving chemical potentials ($z^{I}$, $\wt z$) is:
\be\label{CFTpartition}
 Z_{\CFT_2} = \Tr \left[ e^{2\pi i \tau ( L_0 - {c\over 24} ) -
2\pi i \bar\tau ( \wt L_0 -  {\wt c \over 24})}e^{2\pi i z_I q^I}e^{-2\pi
i \wt z \wt q} \right] \, .
\ee
The dual gravitational theory now has gauge fields $A_{I}, \wt A$ whose boundary values 
are set by the chemical potentials. As mentioned in the introduction, we cannot fix 
the boundary values of both $A_{w}, A_{\wb}$ due to the first order nature of the Chern-Simons 
action which governs the physics at long distances. 

The gravitational path integral should have the form  
\be\label{gravpartition}
Z_{\rm grav} = \int\! [{\cal D} \Phi] ~e^{-\mathcal{S} -{i \over 2\pi}
\int dwd\wb (A^{w}_I J^I_w+ \wt A^{\wb} \wt J_{\wb})}  \equiv 
\int\! [{\cal D} \Phi] ~e^{-\mathcal{S}^{\rm eff}} \, . 
\ee
As discussed in the previous subsection, this is achieved by introducing a boundary term 
in addition to the Chern-Simons action \eqref{Sgauge}. In the classical theory, the currents 
are simply given by the variation of the action with respects to the gauge fields, and one gets 
\eqref{current}. 

To compute the bulk functional integral, we need to evaluate the action for the solutions 
that contribute, including boundary counterterms.
For an on-shell solution around the $\AdS_{3}$ vacuum, one can evaluate the action by the 
following trick  \cite{Balasubramanian:1999re}. One first evaluates the 
variation of the action with respect to the boundary metric $g^{(0)}$ and the 
gauge fields $A^{(0)}, \wt A^{(0)}$:
\be\label{variation}
\delta \mathcal{S}^{\rm eff} =  \int  \! d^2x \sqrt{g^{(0)}}\,\left[
\frac12 T^{ij} \delta g^{(0)}_{ij}+\frac{i}{2\pi} J^{Ii} \delta  A^{(0)}_{Ii}
+\frac{i}{2\pi}\wt J^{i} \delta \wt A^{(0)}_{i} \right]~.
\ee
Using the expressions \eqref{Tgrav}, \eqref{Tgauge}, \eqref{current}, 
reexpressing in the complex coordinates of the boundary metric, and 
integrating the above equation \eqref{variation}, one obtains:
\bea\label{PFgeneral}
\mathcal{S}^{\rm eff}(\tau) & = & - 2 \pi i \t \left(L^{\rm grav}_{0} - \frac{c}{24} \right) + 2 \pi i \bar\t 
\left(\wt L^{\rm grav}_{0} - \frac{\wt c}{24} \right) \cr
&& \; - \frac{i\pi}{2} k \big( \t A_{w}^{2} + \bar \t A_{\wb}^{2} + 2 \bar \t A_{w} A_{\wb} \big) 
+ \frac{i\pi}{2} \wt k \big( \t \wt A_{w}^{2} + \bar \t \wt A_{\wb}^{2} + 2  \t \wt A_{w} \wt A_{\wb} \big)  \, . 
\eea
The first line in the above expression comes from the gravitational part of the action.
The above trick is, in fact, not necessary and one can 
actually evaluate the two-derivative action with appropriate counterterms on the solution 
to get exactly the same result. In our analysis in the following sections, we will go to an 
off-shell theory, and we will evaluate the action on our solutions which have non-trivial 
auxiliary fields turned on. The second line in \eqref{PFgeneral}, on the other hand, comes purely from the 
boundary piece of the gauge field action \eqref{Sgauge}, which is universal since it is the 
boundary term accompanying the universal Chern-Simons action. Since our off-shell solutions 
will asymptote to the on-shell solutions that we have been considering so far, we can 
use the contribution of the boundary terms of the gauge fields exactly as above.

One can use the above expression \eqref{PFgeneral} to find the contribution of the various solutions.
For pure $\AdS_{3}$ and fluctuations around it, one uses  \eqref{PFgeneral} directly, while for the 
black hole and fluctuations around it, one needs to implement the modular transform $\t \to -1/\t$ 
on the various fields of the solution. 
Following this procedure, and correctly identifying \cite{Kraus:2006nb} the boundary gauge fields $A_{I}$ 
in \eqref{gravpartition}, and the chemical potentials $z_{I}$ in \eqref{CFTpartition},  one finds 
the leading term of the action in the high temperature regime as
\be \label{Szzbfinal}
\mathcal{S}^{\rm eff}=-{i\pi k \over 2 \tau}+{i\pi \wt k \over 2\bar\tau} + {2\pi i k z^2 \over
\tau} - {2\pi i \wt k z^2 \over \bar\tau}-{\pi\over \tau_2}   (k z^2+ \wt k \wt z^2)~.
\ee
Using this expression and a saddle point evaluation of the functional integral, 
we find the black hole entropy which is consistent with the Cardy formula. 
To compare with the $\CFT_{2}$ thermal partition function, one needs to 
take into account a subtlety about the relation between the Lagrangian and 
Hamiltonian formulations \cite{Kraus:2006nb}, which amounts to a shift of the 
final two quadratic terms in \eqref{PFgeneral}. 
Bearing this fact in mind, one finds that, indeed the black hole contribution is precisely the leading term 
in the thermal $\CFT_{2}$ partition function.

\subsection{The elliptic genus in supergravity \label{ellgen}}

The elliptic genus is a superconformal index of a $(0,2)$ $\SCFT_2$ defined as 
\bea
\chi(\tau,z_I)= \Tr_{R} \left[ (-1)^Fe^{2\pi \i  \tau
(L_0 -c/24)} e^{-2\pi i \bar\tau (\Lt_0-\wt{c}/24)} e^{2\pi i z_I q^I}\right] \, , 
\eea
where the trace is taken in the R sector of the $\SCFT_2$ living 
on a torus $T^2_{\rm bdry}$ of modular parameter~$\tau$.  The $\SCFT_2$ has a set of left-moving 
$\U(1)$ charges $q^{I}$ with corresponding chemical potentials $z_{I}$. 
Due to the pairing of massive modes due to $(-1)^{F}$, only the ground 
states, with $\Lt_0-\wt{c}/24=0$, contribute in the right-moving sector, 
so the elliptic genus does not depend on $\bar\tau$.    
On the other hand, all left-moving states can contribute.  The elliptic genus is invariant under
smooth deformations of the $\SCFT_2$.  This follow from the quantization of the
charges and of $L_0 - \Lt_0$, together with the fact that only right-moving
ground states contribute.

We are interested in the contributions to the elliptic genus written as a bulk functional integral.
The trace over the R sector with $(-1)^{F}$ insertion dictates that the fermions must be periodic 
around both the circles. The elliptic genus can thus be expressed as  the functional integral
\be
\chi(\tau,z_I)=\left\langle\exp\left[-\frac i{2\pi}z_{I} \int  J^{I} \right]\right\rangle_{++} \, . 
\ee
Naively, it may seem that the elliptic genus receives contributions from the extremal 
BTZ black hole similar to how the thermal BTZ black hole contributes to the 
thermal partition function discussed in the previous section. Indeed, the extremal BTZ 
black hole has $\wt L_{0} -\wt c/24=0$, as mentioned below \eqref{Tgrav}. 

However, as mentioned in the introduction, 
the extremal BTZ black hole solution does not extremize the variational problem 
with a regular boundary torus $T^2_{\rm bdry}$ of complex structure $\t$ (with $\bar \tau = \tau^{*}$ as we have).
The extremal BTZ black hole has a boundary complex structure with $\tau$ finite and $\bar \tau =0$,
as one sees from \eqref{tauinf}. 
For the regular torus, the solution of the equations of motion which fills it in 
is the maximally non-extremal BTZ black hole with $J=0$, $\Rightarrow L_{0} = \wt L_{0}$  
-- even if the action is supersymmetric.  
The other problem with the extremal black hole geometry is, 
as discussed in the introduction, that the contractible circle forces the fermions to have 
anti-periodic boundary conditions.

Both these problems can be resolved if we turn on a chemical potential $\wt z=1/2$ on top 
of the black hole geometry. 
The corresponding bulk operation is to turn on a constant gauge field at the boundary. 
Since all the fermions are charged under this gauge field, the periodicity is effectively changed. 
Moreover, we see from the discussion around 
\eqref{Szzbfinal} that the appropriate partition 
function becomes purely holomorphic, and it becomes exactly what we expect from the 
$\t \to 0$ expansion of the elliptic genus. 
This operation is well-known in the boundary theory, which is called spectral flow. 
This is what we turn to next. 

\vspace{0.2cm}

\ndt \emph{Spectral flow and an associated problem}

The global vacuum of the theory is the NS sector ground state with anti-periodic boundary conditions, 
whose bulk dual geometry is the global $\AdS_{3}$ \eqref{ads3sol}. 
Our real interest is to study the elliptic genus which is a trace in the R sector. 
To go from the NS to the R sector, one has to 
turn on a constant boundary gauge field (or, equivalently, the potential $z$), 
in the boundary theory. This opearation is known as \emph{spectral flow}.

If we find a bulk dual to the spectral flow operation, we would have effectively have 
solved the problem of finding bulk geometries which contribute to the elliptic genus --
we start from the NS sector vacuum (or the more generally, a chiral primary configuration), 
and flow to the corresponding configuration in the R sector. We can then do a modular 
transformation to find the analog of the high temperature partition function. The bulk dual 
of this operation clearly involves turning on constant gauge fields at infinity. 
  
The problem, as mentioned in the introduction, is that a flat gauge field configuration with 
a non-zero winding around the torus, which is well-defined in the boundary theory, cannot be 
extended into the bulk smoothly. By Stokes theorem, the boundary value of $\oint\wt  A$ around a 
contractible circle measures the flux  enclosed within this circle. If we insist 
that the angular component of the gauge field be constant everywhere, then we encounter 
a delta function singularity at the origin. 
One can imagine deforming the singular flux string configuration to a lump of flux 
concentrated at the center of $\AdS_{3}$, such that the gauge field asymptotes to the 
required value at the boundary. This will solve the smoothness problem, but the 
non-zero flux in the bulk of $\AdS_{3}$ will backreact and destroy the background 
geometry. With no further change in the metric, this will not obey the equations of motion,
and will therefore no longer be a saddle point. 

In our investigations presented below, we find that, in fact, there are no supersymmetric 
configurations of this type that solve the equations of motion. This seems to point to a dead end. 
However, there could be \emph{off-shell configurations} which contribute to the functional integral. 
In the absence of supersymmetry, this would be an impossible problem to solve, since we do 
not have a guiding principle for which class of configurations to consider. 
In our case, we use the technique of supersymmetric localization, which tells us to look for 
the (measure zero) subspace of gravitational configurations which are annihilated by 
a certain supercharge in the off-shell theory. 
This is what we shall do this in the following sections.

\section{Localization of the $\AdS$ functional integral \label{localization}}

In this section, we shall discuss the supersymmetry algebra on $\AdS_{3}$. 
We shall then discuss the  formalism of supersymmetric localization. 
Applying this formalism to the $\AdS_{3}$ functional integral, we shall set up  
the localization equations that we need to solve.

\subsection{$\AdS_{3}$ superalgebra}

The bosonic isometry algebra of an $\AdS_{d+1}$ space is the conformal algebra $\SO(d,2)$ 
in $d$ dimensions. For $d=2$, the conformal algebra is an infinite dimensional algebra, 
which spilts up into separate holomorphic and anti-holomorphic pieces. 
One has a number $\CN$ of holomorphic spin 3/2 supercurrents 
and holomorphic spin one $R$-symmetry currents in addition to the  spin 
two stress tensors. The number of currents can be $\CN=$1, 2, or 4 in order to preserve the linear
local superconformal symmetry\footnote{Superconformal theories with $\CN=8$
supersymmetry have been discussed in the $\AdS_{3}$ context in \cite{Dabholkar:2007gp,Kraus:2007vu}. 
In this case, the global supersymmetry can be extended to a current algebra at the cost of 
introducing non-linearity and non-unitarity.}. 
For the $\CN=2$ superconformal algebra, the $R$-symmetry group is $\U(1)$, and 
for the $\CN=4$ algebra, the $R$-symmetry group is $\SU(2)$. 
There is a corresponding structure on the anti-holomorphic side with $\overline{\CN}$
supercurrents. 

In this paper, we shall discuss theories with $(\CN, \overline \CN)=(0,4)$ supersymmetry. 
These could arise as the near-horizon limits of M5-branes wrapping a Calabi-Yau geometry 
in M-theory \cite{Maldacena:1997de}, but this will not be important for us here. There are four supercurrents 
$\wt G^{i\a}(\zb)$ where the index $i$ is in a doublet of the $\SU(2)$ $R$-symmetry (generated 
by the currents $\wt J^{a}(\zb)$, $a=1,2,3$), and the 
index $\a$ is in a doublet of an outer automorphism $\SU(2)'$ symmetry,
which shall later be identified with the $R$-symmetry of the $\CN=2$ supergravity in five dimensions. 
The algebra is as follows: 
\bea \label{scalgebra}
\lbrack \wt L_m,\wt L_n\rbrack&=& (m-n) \wt L_{m+n} + \frac{\tilde c}{12} m (m^2-1) \,\d_{m+n,0} \; , \cr
\lbrack \wt L_n,\wt G_r^{i\a} \rbrack &=& \left(\tfrac{n}{2} - r\right)\, \wt G_{r+n}^{i\a} \; , \cr
\{\wt G^{i+}_{r},\wt G^{j-}_{s}\} &=& 2\,\d^{ij} \wt L_{r+s} + (r-s) \s^a_{ij}\,\wt J_{r+s}^a +\frac{\tilde c}{3}\, \left(r^2 -\tfrac{1}{4}\right)\, \d_{r+s,0}\,\d^{ij} \; , \cr
\{\wt G^{i+}_{r},\wt G^{j+}_{s}\} &=&\{\wt G^{i-}_{r},\wt G^{j-}_{s}\} =0 \; , \cr
\lbrack \wt L_n,\wt J_m^a \rbrack &=& -m \,\wt J_{m+n}^a \; , \cr
\lbrack \wt J_n^a,\wt G_r^{i+} \rbrack &=&\s^a_{ij}\, \wt G_{r+n}^{j+} \, , 
\qquad 
\lbrack \wt J_n^a,\wt G_r^{i-} \rbrack = - \wt G_{r+n}^{j-}\,\s^a_{ji} \; , \cr
\lbrack \wt J_m^a,\wt J_n^b \rbrack&=&-2i\e^{abc}\,\wt J^c_{m+n}+ \frac{\tilde c}{3} m \,\d_{m+n,0}\,\d^{ab} \; .
\eea

In the NS sector, the supercharges are half-integer moded, and the eight supercharges 
$\wt G^{i\a}_{r}$, $r= \pm \half$, along with $\wt L_{0,\pm1}$ and the $\SU(2)$ charges  
$\wt J^{a}_{0}, a =1,2,3$ make up the supergroup $\SU(1, 1|2)$. 
The $\SU(2)$ $R$-symmetry algebra is represented geometrically in a minimal way as 
the rotations of an $S^{2}$. 
In the following sections, we shall consider 5d supergravity theories with asymptotically 
$\AdS_{3} \times S^{2}$ boundary conditions. The global (NS) vacuum will be pure 
$\AdS_{3} \times S^{2}$, this is annihilated by all the eight supercharges of the $\SU(1, 1|2)$, 
this solution will be presented in \S\ref{maximalSUSY}. 

In theories with $\CN=(0,4)$ supersymmetry, the $\wt J^{3}$ component of the right-moving 
$\SU(2)$ $R$-symmetry algebra is related to the fermion number as $F = 2 \wt J^{3}_{0}$. 
By turning on a constant gauge field dual to this charge, we can spectrally flow from the 
NS to the R sector. The spectral flow relates the $\wt  L_{0}$ generators in the two sectors as 
\be\label{L0RL0NS}
\wt L_{0}^{R} = \wt L_{0}^{NS} -\wt  J^{3 NS}_{0}/2 \, .
\ee

In order to apply the formalism 
of localization, we will need to find a subalgebra generated by one real supercharge $Q$ which obeys 
\be\label{Q2L0}
Q^{2} =\wt L_{0}^{R} \, .
\ee
Since the supercharges are integer moded in the R sector, it is clear 
from the algebra \eqref{scalgebra}, and the spectral flow \eqref{L0RL0NS}, that all the zero modes 
$\wt G^{i\a}_{0}$ obey the equation \eqref{Q2L0}. 
The right hand side of \eqref{Q2L0} is manifestly compact, since $\wt L^{NS}_{0}$ is a compact generator 
of the Virasoro algebra acting on $\AdS_{3}$ \eqref{BHVir}, and $\wt  J^{3NS}_{0}$ is the generator 
of rotation on $S^{2}$.

\subsection{Localization}

Let us recall a few facts about the localization of integrals over supermanifolds \cite{Duistermaat:1982vw,Witten:1988ze,Witten:1991mk,Witten:1991zz,Schwarz:1995dg,Zaboronsky:1996qn}.
Consider a supermanifold $\mathcal{M}$ with an integration measure $d\mu$. Let $Q$ be an odd (fermionic) vector field on this manifold under which the measure is invariant and which squares to a compact bosonic symmetry $H$. 
Consider the integral
\begin{equation}
I := \int_{\mathcal{M}} d\mu   \, e^{\mathcal{- S}} .
\end{equation}
where $\mathcal S$ is a $Q$-invariant action. 
To evaluate this integral using  localization, one first deforms the  integral to
\begin{equation}
I (\lambda)  = \int_{\mathcal{M}} d\mu  \, e^{-\mathcal{S}  - \lambda QV} \ , 
\end{equation}
where $V$ is a fermionic, $H$-invariant function which means  $Q^{2} V = 0$ and  $Q V$ is $Q$-exact.  It is easy to see that the derivative of $I(\lambda)$ with respect to $\lambda$ vanishes using the $Q$-invariance of $h$, $\mathcal S$, and the $d\mu$. 
One can thus perform the integral $I(\lambda)$  in the limit of large  $\lambda$ instead of at $\lambda=0$. 
In this limit, the semiclassical evaluation of the functional integral becomes exact and in particular it  localizes onto the  critical points of the functional $\mathcal{S}^{Q} := QV$. 
One can choose 
\begin{equation}\label{locV}
V = (Q\Psi, \Psi) \ ,
\end{equation}
 where $\Psi$ are the fermionic coordinates with some positive definite inner product  defined on the fermions.
In this case, the bosonic part of  ${\mathcal S}^Q$ can be written as a perfect square $(Q\Psi, Q\Psi)$, and hence critical points of ${\mathcal S}^Q$ are the same as the critical points of $Q$ which we refer to as the localizing solutions. Let us denote this set of critical points of $Q$ by $\mathcal{M}_{Q}$. 
The reasoning above shows that the integral over the supermanifold $\mathcal{M}$ localizes to an integral over the submanifold $\mathcal{M}_{Q}$. 
In the large $\lambda$ limit, the integration for directions transverse can be performed exactly in the saddle point evaluation. One is then left with an integral over the submanifold $\mathcal{M}_{Q}$
\begin{equation}
I = \int_{\mathcal{M}_{Q}} d\mu_{Q} \, e^{-\mathcal{S}} \, ,
\end{equation}
with a measure $d\mu_{Q}$ induced on the submanifold by the original measure. 

In our case, $\mathcal{M}$ is the field space of off-shell supergravity, $\mathcal{S}$ is the off-shell supergravity action with appropriate boundary terms. We need to pick a subalgebra of the 
full supersymmetry algebra discussed  above, whose bosonic generator is compact. 
As discussed above, any $Q$ which is a zero mode of one of the supercurrents $\wt G^{i\a}$ 
squares to the compact generator $\wt L_{0}-\wt J^{3}_{0}$. 
The localizing Lagrangian is then defined by 
\begin{equation}\label{loclag}
\CL^{Q} := QV \quad {\rm with} \quad V := (Q \Psi, \Psi) \, ,
\end{equation}
where $\Psi$ refers to all fermions in the theory. The localizing action is then defined by
\begin{equation}
\mathcal{S}^{Q} = \int d^{4} x \sqrt{ g} \, \CL^{Q} \, .
\end{equation}
The localization equations that follow from this action  are
\begin{equation}
Q \Psi = 0 \, . 
\end{equation}
These are the equations that we need  to solve subject to the 
$\AdS_{3}$ boundary conditions, we shall find $Q$ as a particular linear 
combination of $\wt G^{i\a}_{0}$ in the following sections.

\section{$5d$ off-shell supergravity and the $\AdS_{3} \times S^{2}$ solution \label{5dreview}}

Since localization is employed at the level of the functional integral and not just at the level
of a classical action, it is important to use an off-shell formulation of supergravity.
A convenient method for dealing with off-shell formulations of supergravity theories is 
provided by the superconformal multiplet calculus. This calculus was originally constructed 
for the 4d $\CN=2$ supergravity  \cite{deWit:1979ug,deWit:1980tn,deWit:1984pk,deWit:1984px}.  
For 5d supergravity, the conformal supergravity approach was developed relatively recently by 
several groups \cite{Bergshoeff:2001hc,Fujita:2001kv,Bergshoeff:2004kh,Hanaki:2006pj,deWit:2009de},
and these results were exploited further in \cite{Kraus:2005vz,Castro:2007sd,Castro:2007hc,Castro:2008ne}. 
Since this section is meant to set the stage for our computations in the following sections, 
we shall give a very brief summary of the subject.  The interested reader is referred to \cite{deWit:2009de} 
for a more detailed recent treatment, whose notations and conventions we shall follow here. 

The main idea is that the Poincar\'e algebra is extended to the 
superconformal algebra to obtain an off-shell version of the Poincar\'e supergravity. For the 5d $\CN=2$ supergravity, 
the superconformal algebra is given by the exceptional superalgebra $F^2(4)$ whose bosonic part is 
$\SO(2,5)\times \SU(2)'$. Extending conformal supergravity to a gauge theory of 
this superalgebra provides an irreducible off-shell realization of the gravity and matter multiplets. 
Then, by imposing constraints, the gauge theory is identified as a gravity theory. Upon gauge fixing 
the extra superconformal symmetries, one obtains the Poincar\'e supergravity. In this formalism, 
the supersymmetry transformation laws do not depend on the form of the action, and are completely fixed by the 
superconformal algebra.

In \S\ref{Multiplets}, 
we list the multiplets of the  5d  $\CN=2$ superconformal theory that will enter the theories we consider, 
and discuss the supersymmetry transformations and the invariant action of the theory. 
In \S\ref{maximalSUSY}, we review the maximally supersymmetric pure $\AdS_3\times S^2$ configuration.
This will serve as a guide to the new off-shell solutions that we shall find in the next section.
The conventions are spelled out in appendix \S\ref{conventions}. 
More details of the various multiplets
including the full supersymmetry transformation rules can be found in \cite{deWit:2009de},
here we shall only reproduce the formulas that are relevant for the calculations in the 
following sections.

\subsection{Superconformal multiplets and superconformal action}\label{Multiplets}

We consider the 5d $\CN=2$ supergravity coupled to an arbitrary number 
of vector multiplets. We need one other compensating multiplet to eliminate 
the extra degree of freedom, we shall choose this to be a hypermultplet. 
The theory has an $R$-symmetry $\SU(2)'$, under which 
all the fermionic fields are doublets. 
We use the notation that greek indices $(\mu,\nu,\dots)$ indicate the curved spacetime, 
latin indices $(a,b,\dots)$ indicate the flat tangent space indices, and $(i,j, \dots)$ denote 
the $\SU(2)'$ indices. The $\SU(2)'$ indices are raised and lowered  by complex conjugation. 
All the fermionic fields are represented by symplectic-Majorana spinors.

We now describe the field content of the various multiplets. 
We shall use the term ``auxiliary field'' below to mean a field which becomes an auxiliary field in the 
gauge fixed theory. 
\begin{itemize}
\item{Weyl multiplet} \\
\begin{equation}\label{Weyl}
{\bf W}=(e^{a}_\mu,\omega^{ab}_\mu,b_\mu,f^a_\mu,\psi^i_\mu,\phi^i_\mu,V^{ij}_{\mu},T_{ab},D,\chi^i).
\end{equation}
The fields $(e_{\mu}^{a}, w_{\mu}^{ab})$ are the gauge fields for translations (vielbien) and Lorentz 
transformations (spin connection);
$\psi_{\mu}^{i}, \phi_{\mu}^{i}$ are the gauge fields for $Q$-supersymmetries and the  conformal $S$-supersymmetries; 
$(b_{\mu}, f_{\mu}^{a})$ are the gauge fields for dilatations and the special conformal transformations; 
and $V^{ij}_{\mu}$ are the gauge fields for the $\SU(2)'$ $R$-symmetries. 
The $\SU(2)'$ doublet of spinors $\chi^{i}$, the antisymmetric two-form field 
$T_{ab}$ and the real scalar field $D$ are auxiliary fields, 
some of these will play a non-trivial role later.

\item Vector multiplet \\ 
\begin{equation}\label{vector}
{\bf V}=(\sigma^I,W^I_\mu,Y^{I}_{ij},\Omega^{I}_i) \, .
\end{equation}
Here, $W^I_\mu$ are the gauge fields, and $\sigma^I$ are real scalar fields,
$\Omega^I_i$ are the gaugini, and  
$Y^I_{ij}$ is a triplet of  auxiliary fields. The index $I$ labels the
generators of the gauge group $G$. Here, we consider $G$ to be $n_V+1$ copies of $\U(1)$. 
The field strength is given by
\begin{equation}
  \label{eq:W-field-strength}
 \hat F^I_{\mu\nu}= \partial_\mu W^I_\nu - \partial_\nu W^I_\mu -
\bar\Omega^I_i\gamma_{[\mu} \psi_{\nu]}{}^i +\tfrac{1}{2}\i 
\sigma^I\,\bar\psi_{[\mu i} \psi_{\nu]}{}^i \,.
\end{equation}

\item Hypermultiplet \\
The components of ${\bf H}$, hypermultiplets, are
\begin{equation}\label{hyper}
{\bf H}=({\cal A}^\alpha_i,\zeta^\alpha,{\cal F}^i_\alpha),
\end{equation}
where the indices $\alpha=1\cdots 2r$ (in our case $r=1$) label the fundamental representation of $\USp(2r)$. 
The hypermultiplet scalars $\CA_i^\a (\phi)$ can be realized as a section of $\USp(2r)\times \SU(2)$ 
bundle over the hyper-K\"ahler cone whose coordinates are locally written in terms of $\phi$. 
The information on the target-space metric is contained in the so-called
hyper-K\"ahler potential,
\begin{equation}
  \label{eq:hyperkahler-pot}
  \varepsilon_{ij} \,\chi  = \Omega_{\alpha\beta} \,{\cal A}_i{}^\alpha
  {\cal A}_j{}^\beta \, , 
\end{equation}
where $\Omega$ is the skew-symmetric symplectic USp(2N) invariant tensor.

\end{itemize}

Below, we shall need the covariant derivative $\mathcal{D}_\mu$ which is covariant with respect 
to all the bosonic gauge symmetries with the exception of the conformal boosts. Acting on the spinors,
it has the form:
\begin{equation}
  \label{eq:D-epsilon}
\mathcal{D}_{\mu} \epsilon^i = \big( \partial_\mu - \tfrac{1}{4}
\omega_\mu{}^{cd} \, \gamma_{cd} + \tfrac1{2} \, b_\mu\big)
\epsilon^i + \tfrac1{2} \,{V}_{\mu j}{}^i \, \epsilon^j  \, .
\end{equation}
On the hypermultiplets, the covariant derivative takes the form 
\be
{\cal D}_\mu \CA_i{}^\a =  \partial_\mu \CA_i{}^\a -
\tfrac32b_\mu \CA_i{}^\a +\tfrac12V_{\mu i}{}^j\CA_j{}^\a 
+\partial_\mu\phi^A\Gamma^{\ \a}_{A \ \beta}\CA_i^{~\beta} \; ,
\ee
where $\Gamma_A{}^\alpha{}_\beta$ is the $\USp(2r)$ connection associated with rotations 
of the fermions.

Before solving the BPS equations, we will fix a number of gauge conditions. 
We first fix the K-gauge by setting the dilatational gauge field to zero, $b_\mu=0$. 
Then, we identify ${\cal A}_\a^i={\rm const} \times \delta_\a^i$ to gauge-fix the $\SU(2)'$ symmetry. 
In addition, we set the fields $V_\mu^{ij}$ to their on-shell values $V_\mu^{ij}=0$,
and all the fermion backgrounds to zero. 
Now we can fix the $S$-supersymmetry generated by $\eta_{i}$ by solving the purely bosonic 
part of the BPS condition for the hypermultiplet fermion 
\begin{equation}\label{eaaa}
\delta\zeta^\alpha = -\tfrac12 \mathrm{i}{\CDslash}
  \CA_i{}^\alpha\epsilon^i 
  + \tfrac3{2} \CA_i{}^\alpha\eta^i =0~.
\end{equation}
For the field configurations $b_\mu=0, \ V_\mu^{ij}=0$,  the covariant derivative 
\eqref{eq:D-epsilon} becomes a regular partial derivative, which vanishes for the configuration 
${\cal A}_\a^i= {\rm const} \times\delta_\a^i$. We therefore have $\eta^i=0$ for the solution of  \eqref{eaaa}.

With this gauge choice, and considering configurations where there are no fermion bilinears 
in the background, the supersymmetry variations of the fermionic fields are:
\bea\label{gsusy}
\delta \psi^i_\mu&=&\left[{\cal D}_\mu+\tfrac1{4}\i \,
  T_{ab}( 3\,\gamma^{ab}\gamma_\mu-\gamma_\mu\gamma^{ab})\right]\epsilon^i=0~, \label{keq}\\
\delta \Omega^{Ii}&=&
 - \tfrac14 (\hat{F}^I_{ab}- 4\,\sigma^I T_{ab}) \gamma^{ab} \epsilon^i
  -\tfrac1{2}\i  \CDslash \sigma^I\epsilon^i -\varepsilon_{jk}\,
  Y^{Iij} \epsilon^k =0   
  ~, \label{gaugino}\\ 
\delta \chi^i&=&  \tfrac 14 \epsilon^i D 
  + \tfrac3{128}\i (3\, \gamma^{ab} \CDslash
  +\CDslash\gamma^{ab})T_{ab} \, \epsilon^i 
  -\tfrac 3{32} T_{ab}T_{cd}\gamma^{abcd}\epsilon^i =0 \label{chi}  \,,
  \eea
where we have the covariant derivative $\CD_\mu=\partial_\mu -\frac14\omega_{\mu}^{ab} \gamma_{ab}$ 
due to the gauge-fixing. These are the equations we will now solve in the rest of the paper.

The two-derivative bosonic Lagrangian is:
\begin{equation}
  \label{eq:L-tot}
  \mathcal{L} = \mathcal{L}_\mathrm{VVV}+\mathcal{L}_\mathrm{hyper} \, .
\end{equation}
The first piece is the Lagrangian cubic in vector multiplets: 
\begin{eqnarray}
  \label{eq:VVV}
  8\pi^2\mathcal{L}_\mathrm{VVV} &=&{}
  3\,C_{IJK} \sigma^I\Big[\tfrac12 \mathcal{D}_\mu\sigma^J
  \,\mathcal{D}^\mu\sigma^K 
  + \tfrac14 F_{\mu\nu}{}^J F^{\mu\nu K} - Y_{ij}{}^J  Y^{ijK }
  -3\,\sigma^J F_{\mu\nu}{}^K  T^{\mu\nu} \Big] \nonumber \\
  &&{}
   + \tfrac\i 8 C_{IJK}\,e^{-1}
  \varepsilon^{\mu\nu\rho\sigma\tau} W_\mu{}^I 
  F_{\nu\rho}{}^J F_{\sigma\tau}{}^K 
  - C(\sigma) \Big[\tfrac18 \mathcal{R} - 4\,D - \tfrac{39}2
  T^2\Big]\,, 
\end{eqnarray}
where  $C(\sigma)=C_{IJK}\sigma^I\sigma^J\sigma^K$. The second piece is the 
Lagrangian for the hypermultiplets (recall that we have only one hypermultiplet which acts 
as the compensating multiplet):
\begin{equation}
  \label{eq:lagr-hypers}
  8\pi^2\mathcal{L}_\mathrm{hyper} =  -\tfrac12
  \Omega_{\alpha\beta}\, \varepsilon^{ij} 
  {\cal D}_\mu \CA_i{}^\alpha\, {\cal D}^\mu
  \CA_j{}^{\beta}+ \chi\Big[\tfrac{3}{16}\mathcal{R} 
  + 2\, D  + \tfrac3{4} T^2 \Big] \, .
\end{equation}
The constant in the gauge fixing condition for ${\cal A}_\a^i={\rm const}\times \delta_\a^i$ is fixed in such a way that the hyper-K\"ahler potential $\chi$ satisfies   $\chi = -2 C(\s)$,
to ensure that there is no tadpole for the scalar $D$.

We note that we have gauge fixed all the extra symmetries of the superconformal 
supergravity, compared to the Poincar\'e supergravity \emph{except} the local scaling
or dilatation symmetry. Keeping this symmetry unfixed has the advantage of keeping 
the symplectic symmetry acting on the vector multiplets manifest. After finding solutions
of the BPS equations, we will need to fix this remaining dilatation symmetry.

\subsection{The maximally supersymmetric $\AdS_3\times S^2$ solution \label{maximalSUSY}}

The above theory has a maximally supersymmetric pure $\AdS_3\times S^2$ on-shell solution which we 
shall now review. 
This configuration can be interpreted as the magnetic attractor solution since $\AdS_3\times S^2$ 
is the near horizon geometry of a magnetic string 
\cite{Gauntlett:2002nw,Castro:2007sd,Castro:2007hc,Castro:2008ne}. 
This solution is important because it is the bulk dual of the NS sector ground state of 
the boundary $\SCFT_2$, our review will also serve to illustrate the workings of the theory 
as a warmup for the next section. 

The maximally supersymmetric configurations also satisfy the equations of motion 
of the theory, and all the auxiliary fields are set to their on-shell values. 
Since we have maximal supersymmetry, there are no projection conditions 
on the Killing spinor $\e$, and therefore terms in the BPS equations 
with different structures of the gamma matrices should independently vanish. 
The gaugino variation equation \eqref{gaugino} then leads to 
\be
\sigma^I = {\rm constant} \, , \qquad F^I =  4\sigma^I T \, , \qquad  Y^I_{ij}=0 \; .
\label{sigFI}
\ee
Here we can see the characteristic feature of the attractor solutions that the scalar field takes 
constant values in terms of the gauge field charges. 
With this identification, we see that the gravitino variation equation \eqref{keq} becomes 
the usual Killing spinor equation in on-shell treatments.

Next, let us look at the equation \eqref{chi}. 
Imposing vanishing of the independent tensor structure of the $\gamma$-matrices gives:
\be\label{gD}
D =0~,
\ee
and equations which the auxiliary two-form $T$ must satisfy:
\bea\label{gveom}
\varepsilon^{abcde} {\cal D}_a T_{bc} &=& 0~, \label{Bianchi} \\
\i{\cal D}^b T_{ba}  -\varepsilon_{abcde} T^{bc} T^{de}&=&0~. \label{eom}
\eea
The first equation \eqref{Bianchi} is the Bianchi identity and
the second \eqref{eom} is equivalent to the equation of motion for the gauge fields, as 
consistent with the identification of the two as above \eqref{sigFI}. 

Finally, we proceed to consider the Killing spinor equations \eqref{keq}. The geometries which admit 
full Killing spinors are classified in \cite{Gauntlett:2002nw}. Among those, we are interested in 
$\AdS_3\times S^2$ which is interpreted as the near horizon geometry of a magnetic string. 
The metric has the form
 \bea
ds^2=ds^2_{\AdS_3}+ds^2_{S^2}~.
\label{decomposition}
\eea
The gauge field strength $F^I$, (and therefore the auxiliary two-form $T$), 
is proportional to the volume form of $S^2$, so that the fluxes of the gauge 
fields through the $S^{2}$ are fixed in terms of the magnetic charges $p_I$:
\be\label{}
 p^I=\frac{1}{8\pi}\int_{S^2}F^I ~.
\ee
We are then left with the task of computing the ratio of the radii of $\AdS_3$ to the one of $S^2$, 
and the relation of the constant value of the scalar field $\sigma^I$ to the magnetic charge $p^I$.

The relation between the two radii is determined by an integrability condition for the commutator 
of covariant derivatives acting on the Killing spinor
$ \ell_{\AdS_3}=2\ell_{S^2} \equiv \ell$. We now choose coordinates so that the metric is 
\bea
ds^2_{\AdS_3}&=&\cosh^2\!\! \rho \;dt^2 + \sinh^2 \!\!\rho \;\ell^2d\phi^2+\ell^2d\rho^2 \; , \cr
ds^2_{S^2}&=&\tfrac{\ell^2}4\left[d\theta^2+\sin^2\theta d\psi^2\right] \; .
\label{metric}
\eea
The auxiliary two-form $T$ then takes the form:
\bea\label{gvstwo}
T=\tfrac{\ell}{4}\sin\theta d\theta\wedge d\psi=\tfrac1{\ell}{\rm vol}_{S^2}~, 
 \eea
using which, one finds the relation $p^I=\ell\sigma^I/2$.

With the metric \eqref{metric}, the Killing spinor equations \eqref{keq} are also decomposed into 
the $\AdS_3$ and $S^2$ part
\bea
\AdS_3&:& \ \ \ D_m \e_{\AdS_3}=\tfrac 1{2 \ell}(\sigma_3\otimes\mathbbm{1}) \gamma_m\e_{\AdS_3} \, ,  \cr
S^2&:& \ \ \ \ \ \ D_j\e_{S^2}=\tfrac 1{ \ell}(\sigma_3\otimes\mathbbm{1}) \gamma_j\e_{S^2} \, .
\label{Kspieqn}
\eea
The solutions to these Killing spinor equations can be written as:
\begin{eqnarray}
\label{4}
\e_{\AdS_3}&=&\exp\left[\tfrac 12\rho(\sigma_3\otimes\mathbbm{1})\gamma_3\right]\exp\left[-\tfrac {\i }2 \left(\phi+\i \tfrac t\ell\right)(\sigma_3\otimes\mathbbm{1})\gamma_1\right]\mathbbm{1}\otimes\e^0_{\AdS_3}\ , \cr
\e_{S^2}&=&\exp\left[\tfrac {\i }{2} \theta\gamma_5\right]\exp\left[\tfrac {\i }{2} \psi (\sigma_3\otimes\mathbbm{1})\right]\e^0_{S^2}\otimes\mathbbm{1} \ ,
\label{maxsusykilling}
\end{eqnarray}
where $\e_0$ is a constant spinor \cite{Coussaert:1993jp,Lu:1998nu} and we choose the gamma matrices as in \eqref{gammabasis}. To see that the Killing spinor \eqref{maxsusykilling} solves the equations \eqref{Kspieqn}, we use the Baker-Campbell-Hausdorff formula \eqref{BCH}.

\section{Localizing BPS  configurations \label{localizingsolns}}

In this section, we shall achieve our stated goal, that is to find smooth BPS configurations   
in the off-shell theory, with $(++)$ boundary condition on fermionic fields. 
As reviewed in \S\ref{ads3partfn}, we need non-zero gauge field holonomies around both the asymptotic circles
of the geometry. To this end, we shall consider a geometry which asymptotically is an $S^2$ fibration 
over $\AdS_3$. 
We do not {\it a priori} fix conditions on the interior, except that it be everywhere smooth.
To implement the fibration, we introduce Kaluza-Klein gauge fields $\wt  A_\mu$. 
As discussed in  \S\ref{ads3partfn}, the KK gauge fields  $\wt  A_\mu(r)$ will have a 
non-trivial dependence on the radial coordinate to preserve smoothness of the 
configurations\footnote{This is, of course, not true in the Lorentzian theory, wherein 
sphere fibrations with constant KK gauge fields have been discussed in the context 
of $\AdS_3\times S^2$ \cite{deBoer:2008fk} and $\AdS_3\times S^3$
\cite{David:1999zb,Balasubramanian:2000rt,Maldacena:2000dr,Lunin:2002bj}.}.

We thus reach the following ansatz. We want an asymptotically $\AdS_{3}$ space with an $S^{2}$ fibration 
so that we have fixed non-zero $\oint \wt A$ around the two circles of the asymptotic $\AdS_{3}$. We would 
like that this configuration satisfies the BPS equations of the off-shell supergravity of \S\ref{5dreview} with 
respect to at least one supercharge, and without necessarily imposing the equations of motion of the theory. 

The BPS analysis is not trivial since we do not know the form of the metric {\it a priori}. This type of problem can be systematically solved by using the methods of 
\cite{Gauntlett:2002nw}, but here we shall make use of the solutions solved in \cite{Izquierdo:1994jz} instead. It will be very interesting to fully classify the solution spectrum of the off-shell BPS equations,
and we intend to return to this problem in the future. The authors of \cite{Izquierdo:1994jz} have found 1/2-BPS solutions to  a supergravity coupled to Chern-Simons theory on asymptotically $\AdS_{3}$ spaces.
We shall lift the solutions in \cite{Izquierdo:1994jz} to supersymmetric solutions of the 5d off-shell supergravity. As in \cite{Izquierdo:1994jz}, we begin with general ansatz of the 5d metric 
\bea
ds^2 &=& \left[ f(r)^2 + \left(\tfrac{u(r)}{ r}\right)^2\right] dt^2 +2u(r)dtd\phi + h(r)^2 dr^2 +r^2d\phi^2\cr
&&\hspace{3cm}+\tfrac{\ell^2}{4}\left[d\theta^2+\sin^2\theta\left\{d\psi+\wt  A_t(r)dt+\wt  A_\phi(r)d\phi\right\}^2\right] \; .
\label{metricansatz}
\eea
The first line represents a three dimensional base space which is asymptotically $\AdS_3$, 
and the second line expresses an $S^2$ fibration over the base space with $(\wt A_t(r), \wt A_\phi(r))$ 
being the KK gauge fields. Note that details 
like the choice of vielbeins and the corresponding spin connections are presented 
in appendix \S\ref{KillingDetails}. We would like to solve the supersymmetry variation equations 
\eqref{keq}, \eqref{gaugino}, \eqref{chi} with this metric.

\vspace{0.2cm}

\ndt \emph{Choice of the auxiliary two-form $T$}

The gravitino variation equations \eqref{keq} are:
\be \label{KillSpin}
\left[\p_\mu-\tfrac14\omega_\mu^{ab}\gamma_{ab}+ \tfrac{\i}{4} \,
  T_{ab}( 3\,\gamma^{ab}\gamma_\mu-\gamma_\mu\gamma^{ab}) \right]\e=0 \; .
\ee
The relevant Killing spinor equations in \cite{Izquierdo:1994jz} are truncated versions of 
\eqref{KillSpin}, obtained by setting the derivative terms $\wt A'_\mu$ of the KK gauge fields to zero. 
In this case, the variation of the gravitino is supported (in the language of our set up) by  
the auxiliary two-form $T$ of the form: 
\be 
T^{(0)}=\tfrac{\ell}{4}\sin\theta d\theta\wedge \left[d\psi+\wt A_t(r)dt+\wt A_\phi(r)d\phi\right] \, . 
\label{Toldvalue}
\ee
So far, this is as in the maximally BPS solution \eqref{gvstwo}. 
If we want to lift the 1/2 BPS solutions of \cite{Izquierdo:1994jz}, we need to kill the  terms proportional to the derivatives of the KK gauge fields $\wt A'_\mu$, 
which appear in the above equation due to their presence in the 
spin connections $\omega_\mu^{ab}$  \eqref{spinconnection}. 
This will involve a deviation $\delta T$ of the auxiliary two-form from its value $T^{(0)}$ proportional 
to  $\wt A'_\mu$.

Here, we shall make educated guesses to find the deviation $\delta T$. Since we are interested in localizing solutions preserving even one supercharge, 
we are allowed to impose further projection conditions on the spinors, which allows us 
to make progress. Hence, we begin with imposing the projection condition
\be
(1-\gamma_5)\e=0~.
\label{projcon1}
\ee
This projection condition is inspired by the form  of metric \eqref{metricansatz}, similar 
projections have been chosen based on the isometries of the metric in \cite{Castro:2007sd}. 
Using this projection condition, one can find the deviations
\bea\label{deviation}
\begin{array}{l}
\delta T^{31}  =  2 \delta T^{24}=\i(u\wt A'_\phi-r^2\wt A'_t)\frac{ \ell\sin\theta}{12r^2fh} \; ,  \cr
\delta T^{23}  =  2 \delta T^{14}=\frac{\i\ell \wt A'_\phi\sin\theta}{12rh}~, \cr
\delta T^{ab}  =  0  \ \ {\rm otherwise}\ ,
\end{array}
\eea
such that no term proportional to $\wt A'$ survives 
in the Killing spinor equations \eqref{KillSpin}. We present the method to obtain this configuration for the auxiliary two-form
in  appendix \S\ref{KillingDetails}.

\vspace{0.2cm}

\ndt \emph{Gravitino variations}

With this choice of the auxiliary two-form $T$, the Killing spinor equations \eqref{KillSpin} can be written as
\bea
\left[\partial_\mu+\i B_\mu^a\gamma_a(\sigma_3\otimes\mathbbm1) 
-\tfrac12\wt A_\mu (\cos \theta \gamma^{45}-\i \sin\theta\gamma^4)\right]\e&=&0 \, , 
\qquad \mu=(t,\phi,r) \, , \label{adspart} \\
\left[D_j-\tfrac1\ell (\sigma_3\otimes\mathbbm{1}) \gamma_j\right]\e&=&0 \, , \qquad  j=(\theta,\psi) \, , 
\label{s2part}
\eea
where $B^a=-\frac14\varepsilon^{abc}\omega_{bc} +\frac {\i }{2\ell}e^a, \  (a,b,c\in \{1,2,3\})$ are
\bea
B^1&=&     \left( {u'\over 4rh} +\frac{\i f}{2\ell}\right)dt + {1\over 2h} d\phi \, , \cr
B^2 &=&    \left[ {1\over 4h}\left({2u^2\over r^3  f}-{uu'\over r^2 f}-2f'\right)
+ \frac{\i u}{ 2\ell r}\right]dt  +\left[ {1\over 4fh}\left({2u\over r} -u'\right) + \frac{\i r}{2\ell}\right]
d\phi \, ,  \cr
B^3 &=&   \left[{1\over 4rf}\left(u'-{2u\over r} \right) + \frac{\i h}{2\ell}\right] dr  \, .
\label{B}
\eea
Since the Killing spinor equations are factorized into the $\AdS_{3}$ (base) part \eqref{adspart}, 
and $S^{2}$ (fiber) part  \eqref{s2part}, we factorize the Killing spinor $\e$ in the same manner:
\be
\e=\exp\left[\tfrac {\i }{2} \theta\gamma_5\right]\exp\left[\tfrac {\i }{2} \psi (\sigma_3\otimes\mathbbm{1})\right]\zeta~.
\label{killingform}
\ee
With this ansatz, $\e$ satisfies the $S^2$ components of the Killing spinor equations \eqref{s2part} 
as in the maximally supersymmetric case \eqref{maxsusykilling}.
In this case, we are left to solve the 3d base components \eqref{adspart} of the Killing spinor equations:
\be
\left[\partial_\mu+\i B_\mu^a\gamma_a(\sigma_3\otimes\mathbbm1) -\tfrac {\i }2\wt A_\mu (\sigma_3\otimes\mathbbm1)\right]\zeta=0  \; .
\label{killing}
\ee
This equation \eqref{killing} is almost the same as the Killing spinor equations 
solved in \cite{Izquierdo:1994jz} as we aimed. (See Eq. (2.3) in \cite{Izquierdo:1994jz}.) We present the method to obtain solutions of these equations and detailed calculations 
in  appendix \S\ref{ads3offshellsolns}, here we present the results. The Killing spinors 
are of the form:
\bea
\zeta=a(r)e^{(\frac{\i n}{ 2}\phi +\frac {\i m}2\frac t\ell)(\sigma_3\otimes\mathbbm1)} (1+\Gamma^a b_a)\zeta_0\ , 
\label{Killingsoln}
\eea
where $n, m$ are integers, $\Gamma_a$ are the 3d gamma matrices defined in 
\eqref{3dgamma}, and $\zeta_0$ is a constant spinor. The real function $a(r)$ and the complex functions 
$b_a(r)$ (written explicitly in \eqref{a}, \eqref{b}) are determined in terms of one real function $u(r)$. 
The solutions \eqref{Killingsoln} obeys the projection condition
\be
(1-\Gamma^ab_a)\zeta=0 \ .
\label{projcon2}
\ee
Since in total we impose two projection conditions, \eqref{projcon1} and \eqref{projcon2}, the Killing spinors \eqref{killingform} with \eqref{Killingsoln} are  1/4 BPS solutions.

The metric admitting the Killing spinors \eqref{Killingsoln} 
is\footnote{Note that we have substituted $\a=\i/\ell, \ \b=1$ and redefined $u\to \i u$ in \eqref{5dmetric}, so that 
the function $u(r)$ now takes real values.}:
\bea\label{5d metric}
ds^2 &=& \left[\left(\frac r\ell\right)^2 +\frac {2 u}\ell + 1\right]dt^2 +2\i u dtd\phi
+\tfrac{\left(1 + \frac{\ell u'}{2r}\right)^2 }{\left(\frac r\ell+ {u\over r}\right)^2 +1}dr^2
+r^2d\phi^2 \cr
&& \qquad \qquad \qquad \qquad + \frac{\ell^2}{4}\left[d\theta^2+\sin^2\theta\left\{d\psi+ \wt A_t  \, dt +
\wt A_\phi d\phi\right\}^2\right] \, , 
\eea
where the KK gauge fields are given by
\be\label{gaugefieldsoln}
\wt A_t =  \frac{\i  u'}{ 2 r +\ell u'} +\frac{\i }{\ell}+\frac m\ell \, , \qquad 
\wt A_\phi =  {2 r\over 2 r +\ell u'} +{n} \; .
\ee
The Killing spinors \eqref{Killingsoln} are periodic $(+)$ for even integers $m,n$. 
We will pick $n=m=0$ to get the spinor zero modes as required.
If we pick the function $u(r)$ to approach a constant asymptotically, we will get 
constant KK gauge fields at infinity as required. We shall analyze the solutions 
in more detail in the next section. 

\vspace{0.2cm}

\ndt \emph{Gaugino variations}

Next, we need to solve the gaugino variation equations: 
\be\label{BPS vector}
\left(  - \tfrac14 {F}^I\!\cdot\! \gamma   +\sigma^I\,T\!\cdot\!\gamma -\tfrac1{2}\i  \Dslash \sigma^I \right)\e^i
 -\varepsilon_{jk}\,
  Y^{Iij} \epsilon^k~=0~.
\ee
The auxiliary two-form $T$ is given by  \eqref{Toldvalue}, \eqref{Taux}. 
The field strength of the gauge field $W^{I}$ given by 
\bea
F^I=\ell\sigma^I\sin\theta d\theta\wedge \left[d\psi+\wt A_tdt+\wt A_\phi d\phi\right]-\ell\sigma^I\cos\theta \left[\wt A'_tdr\wedge dt+\wt A'_\phi dr\wedge d\phi\right]
\label{fs}
\eea
solves the BPS condition \eqref{BPS vector} with the auxiliary field taking the value 
\be\label{Yaux}
Y^{Ii}_{\ \ \:  j}=\pm e^{-\i \theta}\frac{\ell\sigma^I \wt A'_\phi}{4 r f h}  \ \delta^i_{~j} \, , 
\ee 
and constant scalar fields $\sigma^I$. 

One can check that the field strength \eqref{fs} satisfies the Bianchi identity $dF^I=0$.
Actually, there is a very natural construction of this field strength which has been discussed in 
\cite{Hansen:2006wu}.  
If one writes the metric \eqref{metricansatz} as
\bea
ds^2 = ds_{\AdS_{3}}^2 +\tfrac{\ell^2}4  (dy^i -
 \wt A^{ij}y^j)(dy^i - \wt A^{ik}y^k)~,
\eea
where $\sum_{i=1}^3 (y^i)^2 =1$ and the one-forms  $\wt A^{ik}$ are the $\SO(3)$ KK gauge fields, then the form \eqref{fs} of the field strength $F^I$ can be written in terms of this coordinate:
\be \label{fseuclid}
F^I = \tfrac{\ell\sigma^I}2  \epsilon_{ijk}(
Dy^i Dy^j- \wt{F}^{ij}  ) y^k \, , 
\ee
where the one form and modified field strength are defined as 
\be
 Dy^i  = dy^i - \wt A^{ij}y^j \, , \qquad \wt{F}^{ij}  = d \wt A^{ij} - \wt A^{ik}  \wt A^{kj} \; .
\ee

We find that the third BPS equation \eqref{chi} is also solved, if the field $D$ takes on a non-trivial 
dictated by the two-form $T_{ab}$ that we have. 
Since the field $D$ drops out of the action \eqref{effective}, the actual value is not important for our purposes. 

\vspace{0.2cm}

One can ask if a first projection condition like \eqref{projcon1} was even necessary, 
{\it i.e.} can one directly lift the solutions \eqref{Killingsoln} to 1/2 BPS solutions
in five dimensions. One can check by substitution that this is not the case, 
and we really have 1/4-BPS solutions.

\section{Analysis of the localizing solution \label{holomorphy}}

In the previous section, we have found smooth Killing spinors \eqref{Killingsoln} which have $(++)$ boundary conditions around both the circles of the boundary torus $T^2_{\rm bdry}$ . These spinors live on the 
asymptotically $\AdS_{3}$ \eqref{5d metric} with the KK gauge fields \eqref{gaugefieldsoln}. 
They depend on one function $u(r)$ which supersymmetry does not fix. 
In this section, we shall analyze the properties of this solution. We first analyze how 
smoothness restricts the form of the geometry, and then compute the action of the 
smooth configuration that we find.

Firstly, since $u(r)$ is a normalizable deformation, we should have 
(with constant $u_{\infty}$): 
\be
u(r) \to  \ell u_{\infty} + {\cal O}(\tfrac1r) \ \ {\rm as} \ \ r\to \infty \, .
\ee 
This means that the values of the gauge fields at infinity are:
\be
\wt A_{\phi}(r) \stackrel{r \to \infty}{\longrightarrow} 1 \, , \qquad 
\wt A_{t}(r) \stackrel{r \to \infty}{\longrightarrow} \i/\ell \; .
\ee

Smoothness of the configurations requires the angular component of a gauge field to vanish 
at the origin of a contractible circle. The form of the metric dictates that $r=0$ should be the origin. 
Choosing $\phi$ to be the contractible circle, we should impose that 
$\wt A_{\phi}(r)$ vanishes at the origin. This is achieved by
\be
u'(r) \stackrel{r \to 0}{\longrightarrow} \CO(1) \; ,
\ee
which gives 
\be
\wt A_{\phi}(r) \stackrel{r \to 0}{\longrightarrow} 0 \, , \qquad 
\wt A_{t}(r) \stackrel{r \to 0}{\longrightarrow} 2 \i/\ell \; .
\ee

It is convenient to make the combinations
\be
\wt A_{w} \equiv \frac12 \big(\wt A_{\phi} + \i \ell \wt A_{t} \big) \, , \qquad 
\wt A_{\bar w} \equiv \frac12 \big( \wt A_{\phi} - \i \ell \wt A_{t} \big) \, , 
\ee 
which have the limiting values:
\bea \label{Awbdryvalues}
\wt A_{w}(r) \stackrel{r \to \infty}{\longrightarrow} 0 \, , && \qquad 
\wt A_{\bar w}(r) \stackrel{r \to \infty}{\longrightarrow} 1  \; ;\cr
\wt A_{w}(r) \stackrel{r \to 0}{\longrightarrow} -1 \, , && \qquad 
\wt A_{\bar w}(r) \stackrel{r \to 0}{\longrightarrow} 1 \; .
\eea
The field $\wt A_{\bar w}$ which everywhere takes the value $1$ is identified with the 
(appropriately normalized) 
right-moving chemical potential $\wt z$ of the boundary $\SCFT_{2}$ which we 
would like to ``turn on'' to have a value 1/2.

We will now analyze the near-horizon region $r \to 0$. In order to zoom in near this 
region, we use the change of coordinates 
\be
r = \lambda \rho \, , \qquad \lambda \to 0 \; , 
\ee
and keep only the terms which survive in the limit \cite{Sen:2007qy}. The near-horizon metric takes the form:
\be\label{near horizon}
ds^{2} = (2\wt u+1) \, dt^{2} + 2 \i \wt  u \, dt \, d\phi +  \frac{\wt u'^{2}}{4\wt  u^{2}} \, d \rho^{2} + 
\frac14 \big[d\theta^{2} + \sin^{2}\theta \, (d \psi + 2 \i \, dt)^{2}  \big] \, .
\ee
The function $\wt u(\rho)\equiv u(\lambda\rho)$ can be thought of as the effective radial coordinate, and we get a 
non-singular geometry at the origin if we choose $\wt  u (\rho)= \rho/2$, in which case we get:
\be\label{very-near-horizon}
ds^{2} = (\rho +1) \left[dt + \frac{\i \rho d\phi}{2(\rho+1)}\right]^{2} + \frac14 \left[\frac{d \rho^{2}}{\rho^{2}} 
+ \frac{\rho^{2}}{ \rho +1} d \phi^{2} \right] + \frac14 \big[d\theta^{2} + \sin^{2}\theta \, (d \psi + 2 \i \, dt)^{2}  \big] \, .
 \ee

Near $\rho \to 0$, we recognize the familiar form of the $\AdS_{2}$ in the $(\rho,\phi)$
coordinates, with a circle $t$ and a two-sphere $(\theta, \psi)$ fibered over it. 
In fact, the three dimensional space \eqref{near horizon} spanned by $(\rho, \phi, t)$ is a locally $\AdS_{3}$ 
space, as can be checked by computing the Ricci scalar.
The near-horizon region is indeed an $\AdS_{3} \times S^{2}$, with a constant KK gauge field. 
In the very-near-horizon region \cite{Strominger:1998yg}, we see that the geometry \eqref{very-near-horizon} is actually that of the supersymmetric 
Euclidean BTZ black hole. 

It is easy to check that, in the near-horizon limit, the field strengths $F^{I}$ take their 
constant on-shell values determined by the magnetic charges  as in \S\ref{maximalSUSY}, and 
that the auxiliary two-form $\delta T_{ab}$  \eqref{deviation} and $Y^{I}_{ij}$ 
 \eqref{Yaux} vanish in the near-horizon region. Note that the function $u(r)$ is fixed to a certain 
value both near the boundary and near the horizon. 
Since our configuration is supersymmetric, 
this implies that  the solution \eqref{very-near-horizon} is actually on-shell. This can also be  
explicitly checked by verifying that \eqref{very-near-horizon}  solves the equations of the 5d 
$\CN=2$ Poincar\'e supergravity.

\vspace{0.4cm}

\ndt \emph{The action of the configurations}

\vspace{0.2cm}

Now that we have found smooth configurations that contribute to the functional integral, 
we want to know how much they contribute. In general, this can be answered by evaluating 
the action on the solutions. For on-shell configurations, we can actually simplify this step 
and deduce the contribution by the methods as discussed in \S\ref{ads3partfn}. 
Here, since we have configurations that do not solve the equations of motion, 
we shall actually have to evaluate the action on the configurations, which is what we turn to next. 

We recall from \S\ref{5dreview} that the fields of the compensating hypermultiplet take the values 
${\cal A}^i_{\alpha}={\rm const}\times \delta^i_\alpha$ and $ \chi= - 2\,C(\sigma)$. With these values, 
the two-derivative Lagrangian  \eqref{eq:L-tot} is 
\bea\label{effective}
8\pi^2{\cal L}&=& 3\,C_{IJK} \sigma^I\Big[\tfrac14 F_{\mu\nu}{}^J F^{\mu\nu K} - Y_{ij}{}^J  Y^{ijK }
  -3\,\sigma^J F_{\mu\nu}{}^K  T^{\mu\nu} \Big] \cr
   &&+ \tfrac\i 8 C_{IJK}\,e^{-1}
  \varepsilon^{\mu\nu\rho\sigma\tau} W_\mu{}^I 
  F_{\nu\rho}{}^J F_{\sigma\tau}{}^K 
  - C(\sigma) \Big[\tfrac12 \mathcal{R} - 18
  T^2\Big]~.
\eea
It is straightforward to compute the various quantities in the Lagrangian 
evaluated on the localizing solution given by the metric \eqref{5d metric}, 
the auxiliary two-form field $T_{ab}$ \eqref{twoformaux}, the gauge fields $W^{I}$ \eqref{fs} and 
the auxiliary fields $Y^{I}_{ij}$ \eqref{Yaux}. 

The bulk action is obtained by integrating this Lagrangian over the five dimensional space. 
This integral is divergent due to the infinite volume of the asymptotically $\AdS_{3}$ space. 
We shall regulate this divergence by imposing a cutoff at large $r=\Lambda$.
Doing so, one finds (the details of this calculation are presented in appendix \S\ref{actionapp}):
\be\label{bulkint1line}
{\cal S}_{\rm bulk} = -\tfrac{C(\sigma)}{8\pi }\left[ \Lambda^2+\ell^2 u_\infty\right] {\rm Area}(T^2_{\rm bdry}) \; , 
\ee
where $T^2_{\rm bdry}$ denotes the boundary torus with coordinates \eqref{wtorus} on which the $(0,4)$ $\SCFT_2$ lives.

As explained in \S\ref{ads3partfn}, we need to include the Gibbons-Hawking term 
written in terms of the extrinsic curvature at the boundary, as well as the boundary counterterm 
to remove the divergence. 
The details of the evaluation of the boundary terms is also presented in appendix \S\ref{actionapp}. 
The results are:
\bea
{\cal S}_{\rm GH}=\tfrac{C(\sigma)}{8\pi^2 }\int_{r=\Lambda} d^4x \sqrt{h}~K = 
\tfrac{C(\sigma)}{8\pi }  {\rm Area}(T^2_{\rm bdry})\left[2\Lambda^2 +\ell^2(1+2 u_\infty)\right]
+{\cal O}(\tfrac1{\Lambda^2}) \, , 
\eea
and 
\be
{\cal S}_{\rm ct}=-\tfrac{C(\sigma)}{8\pi^2 }\int_{r=\Lambda} d^4x \sqrt{h}~\tfrac1\ell 
= -\tfrac{C(\sigma)}{8\pi }  {\rm Area}(T^2_{\rm bdry})\left[\Lambda^2 +\tfrac{\ell^2}2(1+2 u_\infty)\right]+{\cal O}(\tfrac1{\Lambda^2})~.
\ee

In order to express the final result in a useful manner, we need to evaluate 
the overall factor $C(\s)$ is proportional to the inverse of the 5d Newton's constant $G_5$, which in 
turn can be written in terms of the 3d Newton's constant $G_3$ and the volume of the $S^{2}$. 
This in turn is related to the central charge of the $\AdS_{3}$. These relations can be summarize as follows:
\be \label{relconsts}
c=\wt c=6\wt k=\frac{3\ell^3C(\sigma)}2=\frac{3\pi\ell^3}{2G_5}=\frac{3\ell}{2G_3} \, .
\ee
Putting all these calculations together, and taking the cutoff $\Lambda \to \infty$, we obtain 
\be\label{finalSbulk}
{\CS}_{\rm reg} = {\CS}_{\rm bulk}+{\cal S}_{\rm GH}+{\cal S}_{\rm ct}
= -\frac{i\pi c}{12 \ell}(\tau-\bar\tau)~.
\ee
In the on-shell case reviewed in \S\ref{ads3partfn}, the analogous action, {\it i.e.}
the first line of \eqref{PFgeneral} was computed without much calculation. 
Here, we go through a slightly laborious 
process, but eventually get the simple answer \eqref{finalSbulk}.

Finally, we need to consider the boundary term coming from the Chern-Simons term, 
which is subtle, but standard. 
The careful treatment explained in \S\ref{ads3partfn} yields the second line of \eqref{PFgeneral}. 
\be\label{CSbdry}
{\cal S}_{\rm gauge}^{\rm bdry} = - \frac{i\pi}{2} k \big( \t A_{w}^{2} + \bar \t A_{\wb}^{2} + 2 \bar \t A_{w} A_{\wb} \big) 
+ \frac{i\pi}{2} \wt k \big( \t \wt A_{w}^{2} + \bar \t \wt A_{\wb}^{2} + 2  \t \wt A_{w} \wt A_{\wb} \big)  \, . 
\ee
The only Chern-Simons term that we consider here is for the right-moving gauge field $\wt A$, 
whose boundary values are given by \eqref{Awbdryvalues}. 
Adding this in, we find that the $\bar \tau$ dependence completely drops out 
and we are left with 
\be
{\CS}^{\rm eff} =  {\CS}_{\rm reg} + {\cal S}_{\rm gauge}^{\rm bdry} = -\frac{i\pi c \, \tau}{12 \ell} \, , 
\ee 
which is holomorphic in $\t$. 

We would now like to make a few comments. 
Firstly, recalling that the full effective action can be interpreted in the boundary theory as 
\be
\mathcal{S}^{\rm eff}(\tau)  =  - 2 \pi i \t \left(L_{0} - \frac{c}{24} \right) 
+ 2 \pi i \bar\t \left(\wt L_{0} - \frac{\wt c}{24} \right)  \, , 
\ee
we deduce that the action of our localizing configuration is equal to that of the R sector 
ground state, with $\wt L_{0} = \wt c/24$. 
In our solution, the left-movers seem to be in their ground state, this is due to the form of 
the gauge field strength \eqref{fs}. More generally, we will have purely holomorphic excitations coming 
from turning on left-moving potentials $z^{I}$. 
According to our discussion in \S\ref{ads3partfn}, after modular transformation, this is the same as
the action of the supersymmetric BTZ black hole. When we do not turn on any $z^{I}$, 
we get the action of the zero mass BTZ black hole. 

Secondly, from computing the Ricci curvature of the pieces of the near-horizon region \eqref{very-near-horizon},
we see that it is indeed of the form $\AdS_{3} \times S^{2}$. The value of the Ricci scalar 
\eqref{Ricciscvalue} approaches a constant near the origin, and the contribution of the 
auxiliary fields vanish in this region, making the region on-shell. The boundary region is,
of course, also on-shell, but in order to connect these two smoothly, the solution is forced 
to go off-shell, as can be seen from the fact that, in the intermediate, finite $r$ region, 
the auxiliary fields are non-zero. 

Finally, we see that although the values of the Ricci scalar, the field strengths, the auxiliary two-form $T_{ab}$ \eqref{twoformaux} 
and $Y^{I}_{ij}$ \eqref{Yaux} all depend on the function $u(r)$ in the intermediate region, the value of the total action 
is independent of $u(r)$, indicating the presence of a gauge symmetry. Recalling that in our 
analysis of the BPS equations in \S\ref{5dreview}, we gauge fixed all the gauge symmetries 
of the conformal supergravity \emph{except} for the local dilatation symmetry. 
It is very likely that the (spurious) function $u(r)$ is an indication of this symmetry. 
We can verify this by perturbatively fixing $u(r)$ near the horizon 
and near the boundary, we leave a detailed analysis of this issue for the future. We already 
note that the function $u(r)$ factors out of the functional integral by canceling the volume of the 
gauge group of local dilatations. We thus understand the configuration \eqref{5d metric}
as the unique localizing solution 
in the gravitational sector, replacing the role of the BTZ black hole solution in the gravitational 
computations of supersymmetric functional integrals.

\section{Discussion and future directions \label{discussion}}
The main motivation of this work is to compute the exact gravitational functional integral 
dual to the elliptic genus, or more generally, the superconformal index of a field theory. 
We summarize here the main findings of our work: \\
The bosonic configuration \eqref{5d metric} represents an everywhere smooth, asymptotically 
$\AdS_{3}$ space with an $S^{2}$ fibered over the three-dimensional geometry such that 
the KK gauge fields \eqref{gaugefieldsoln} carry a non-trivial flux. The flux is concentrated near 
the origin and dies away towards infinity such that the gauge fields have a constant winding around 
the torus. The near-horizon configuration is that of a supersymmetric BTZ black hole with constant 
flux. The near-horizon region and the asymptotic region are both solutions to the equations of 
motion of the theory. The intermediate region has non-zero auxiliary fields and is therefore off-shell, but 
nevertheless, it is a solution of the off-shell BPS equations.

We would like to reemphasize that our approach is Euclidean, and our localizing solution is complex. 
We do not have anything to say here about which intermediate states propagate in loops in 
the physical theory. This situation is not unusual in Euclidean functional integrals. 
In the full evaluation, we will of course have to do the usual Gibbons-Hawking analytic continuation, 
similar to \cite{Dabholkar:2010uh}.

This is, however, only the first step in this direction, and in order to really compute the 
elliptic genus from gravity, there are many more steps to take on this path:
firstly, although we have argued that our localizing configurations are special, 
we have not shown that they are the only BPS configurations $(++)$ boundary conditions, 
and we need to work out the full set of supersymmetric gravitational fluctuations. 
Perhaps the methods of \cite{Gauntlett:2002nw} would be useful in this regard. 
It would also be nice to have a more complete treatment of the gauge freedom $u(r)$.

Further, we need to find the most general solution in the vector multiplet sector. These 
solutions should provide a holomorphic action to the system. One may envision 
the $\AdS_{3}$ analog of \cite{Dabholkar:2010uh, Dabholkar:2011ec}, where on integrating these fluctuations, 
one finds the polar part of the elliptic genus. 
In this paper, we only included the two-derivative action of the 5d supergravity, but in general, 
of course, we need to include the full effective action which may contain higher derivative terms. 
Taking all these into account, one can now put the black hole Farey tail \cite{Dijkgraaf:2000fq} on a much more rigorous footing. One imagines starting with our solution in this paper, including the full set of fluctuations 
obeying the localization equations, and then summing over the modular transforms to get the full elliptic 
genus\footnote{Here we assume that there is no wall-crossing while going from weak to strong coupling.
In general, there will be other gravitational configurations. One example where these can be dealt 
with exactly is in \cite{DMZ}.}.  
These issues are currently under investigation. 

Another much bigger issue is what is the full off-shell space of gravity, and whether the 5d off-shell 
supergravity (and in particular, its field space) comes from a more fundamental theory like string theory. 
Our attitude in this paper has been that we have an example of a theory which tells us which off-shell 
configurations to integrate over to compute the supersymmetric functional integral. It seems to be 
able to capture a large class of $\CN=2$ compactifications in five dimensions related to $M$ theory 
on a Calabi-Yau three-fold \cite{Maldacena:1997de}. It seems like we can 
generalize the approach to other dimensions as well.
For example, one can also try to extend our analysis to the $(4,4)$ theory, dual to 
$\AdS_3\times S^3$. This needs a six dimensional off-shell formalism. 

A particularly interesting direction is the analogous story in higher 
dimensions. In this regard, note that, from our point of view of the Euclidean functional integral, 
it is not a puzzle, and in fact completely natural, that for $\AdS_{d \ge 4}$, the most general 
superconformal index does not grow fast enough to accommodate the density of states of a 
black hole \cite{Kinney:2005ej,Bhattacharya:2008zy, Kim:2009wb}. 
It will of course, be interesting, if there is a Lorentzian interpretation of this statement, perhaps 
an existence of a pair of fermionic zero modes in the background of the supersymmetric 
$\AdS_{d \ge 4}$ black hole. Even within the Euclidean functional integral formalism, it would be interesting if 
one can compute the known exact expressions for the superconformal indices in $d \ge 3$
$\SCFT$s using localization in the dual $\AdS_{d+1}$ theories. 

We would like to comment that our localized solutions in $\AdS_{3}$ are really (off-shell) fluctuations 
around the vacuum, and there may be similar localized solutions in $\AdS_{d \ge 4}$. Integrating over 
these may reproduce the relevant (slowly-growing) index in the boundary theory. 
The reason we have come close to reproducing the (exponentially-growing) elliptic genus here 
is that the modular transforms of our solutions have the action of a supersymmetric black hole. 
The known boundary calculations in the higher dimensional case suggest that there may be 
no analog of the modular transformation between the ground state and highly excited states.

\section*{Acknowledgements}
The authors would like to thank Atish Dabholkar, Jan de Boer, Bernard de Wit, Jo\~ao Gomes,
Jaume Gomis, Rajesh Gupta, Janet Hung, Gautum Mandal, Shiraz Minwalla, Robert Myers, 
Sandip Trivedi, Erik Verlinde and especially Ashoke Sen for valuable discussions. 
This work was initiated and partially completed during the academic year 2010-2011, 
when both the authors had an affiliation at the Tata Institute of Fundamental Research (TIFR),   
Mumbai. The hospitality of TIFR is duly acknowledged. 
S.~N.~would like to thank the Simons Summer Workshop in Mathematics and Physics 2011 
for its stimulating academic environment and its warm hospitality since he benefited from 
discussions in the workshop. 
The work of S.~M. was supported by the European 
Commission Marie Curie Fellowship under the contract PIIF-GA-2008-220899, and 
is presently supported in part by the ERC Advanced Grant no. 246974,
{\it ``Supersymmetry: a window to non-perturbative physics''}. The work of S.~N. is supported by the ERA Grant ER08-05-174. Research at Perimeter Institute is supported by the Government of Canada through Industry Canada and by the Province of Ontario through the Ministry of Research and Innovation.

\appendix

\section{Conventions\label{conventions}}

We summarize our notational conventions in this appendix. Firstly, we work with the Euclidean signature $(+\cdots +)$. The curved metric is written by $g_{\mu\nu}$ and the flat metric is denoted by $\delta_{ab}$. Greek letters $\mu, \nu, \ldots$ express coordinate indices, and Latin letters
$a,b,\ldots$ represent orthonormal indices where they take values $1,\dots, 5$. 
All the fermionic fields are represented by symplectic-Majorana spinors.

The f\"unfbein is denoted by $e_{\mu}^{\;\;a}$ and we write its inverse 
by $e_a^{\;\;\mu}$. 
They satisfy $e_{\mu}^{\;\;a} e_a^{\;\;\nu} = \delta_{\mu}^{\nu}$.
The curved and flat metrics are related by
\be
g_{\mu \nu} = e_{\mu}^{\;\;a} e_{\nu}^{\;\;b} \delta_{ab}\;, \;\;\;
\delta_{ab} = e_a^{\;\;\mu} e_b^{\;\;\nu} g_{\mu \nu} \;.
\ee	
Curved and flat indices are converted by
\be
V_a = e_a^{\;\;\mu} V_{\mu} \;, \;\;\;
V_{\mu} = e_{\mu}^{\;\;a} V_a \;.
\ee
Curved and flat indices are moved up and down with $g_{\mu \nu}$ and
$\eta_{ab}$ and their inverses $g^{\mu\nu}$ and $\eta^{ab}$,
respectively. We define
\be
e =\det(e_{\mu }{}^a)= \sqrt{ \det(g_{\mu \nu}) } \;.
\ee
The spin connections\footnote{Note that our convention for the spin connections  is different from the most common one in the literature. Hence, the curvature tensors are also different from the conventional definitions so that the Ricci scalar curvatures for $\AdS$ spaces are positive.}    $\omega^a{}_b=\omega_\mu{}^a{}_{ b}dx^\mu$ are defined by	
\bea
\omega^a{}_b&=&-\omega^b{}_a\ ,\cr
de^a&=&\omega^a{}_b\wedge e^b\ .
\eea
The Riemann tensors are defined by
\be
 \mathcal{R}_{\mu\nu}{}^{ab} = 
  \partial_{\mu}\omega_{\nu}{}^{ab} -  \partial_{\mu}\omega_{\nu }{}^{ab} 
  - \omega_{\mu}{}^{ac} \omega_{\nu c}{}^{b} +\omega_{ \mu}{}^{ac} \omega_{\nu c}{}^{b} \ .
\ee
By contraction, one obtains the Ricci tensors
\be
\mathcal{R}_\mu^{\;\;a} = \mathcal{R}_{\mu\nu}{}^{ab} e_b^{\;\;\nu} \ ,
\ee
and the Ricci scalar
\be
\mathcal{R} = \mathcal{R}_\mu^{\;\;a} e_a^{\;\;\mu} \;.
\ee

Next, we shall fix the convention of spinors. The gamma matrices obey
\be
\{\g_a,\g_b\}=2\delta_{ab} \mathbbm{1}~.
\ee
We define the products of the gamma matrices 
\begin{equation}
  \gamma^{\mu_1\cdots\mu_p}= 
  \gamma^{[\mu_1}\gamma^{\mu_2}\cdots\gamma^{\mu_p]} = \frac{1}{p!} 
\left( \gamma^{\mu_1}\ldots \gamma^{\mu_p} \pm \text{cyclic}\right) ~.
\end{equation}
Then the products of the gamma matrices satisfy
 \bea\label{gammaalg}
  \gamma^{ab}\gamma^{cd} & = & - (\delta^{ac}
\delta^{bd}-\delta^{ad}\delta^{bc}) -(\gamma^{ac}
\delta^{bd}-\gamma^{bc}\delta^{ad} + \gamma^{bd}
\delta^{ac}-\gamma^{ad}\delta^{bc}) +\gamma^{abcd} ~,\cr
\gamma^a\gamma^{bc} & = & \delta^{ab}\gamma^c - \delta^{ac}\gamma^b +
\gamma^{abc}~,\cr \gamma_{abcde}&=&-\mathbbm{1}\;\varepsilon_{abcde}~, 
\eea
where $\varepsilon_{12345}=\varepsilon^{12345}=+1$. To solve the Killing spinor equations explicitly in \S\ref{maximalSUSY} and \S\ref{localizingsolns}, we choose the 5d gamma matrices to be 
\bea\label{gammabasis}
\AdS_3&&\left\{\begin{array}{l}
\gamma_1 = \sigma_3 \otimes \sigma_1\ , \cr
\gamma_2 = \sigma_3 \otimes \sigma_2\ , \cr
\gamma_{3} = \sigma_3\otimes\sigma_3 \ ,
\end{array}\right . \cr
S^2&&\left\{\begin{array}{l}
\gamma_4 =\sigma_1 \otimes \mathbbm{1}\ ,\cr
\gamma_5 = \sigma_2 \otimes \mathbbm{1} \ .
\end{array}\right .
\eea
Note that, with this base choice of the gamma matrices, they satisfy $(\g^a)^\dagger=\g^a$. 

In the manipulations of the gamma matrices, we use the Baker-Campbell-Hausdorff formula 
\be
e^{\tfrac{\i}2\theta\, X}\, Y\, e^{-\tfrac{\i}2\theta\, X} = Y\cos\theta\,  -
Z\sin\theta \, ,
\label{BCH}
\ee
for matrices $X$, $Y$ and $Z$, such that ${[} X, Y{]}= 2\i Z$, and
${[} X,Z{]} = -2\i Y$.

To fix the convention of spinors, we introduce the charge conjugation matrix $\SC$ which satisfies
\begin{eqnarray} 
  \label{eq:spinor-conv} 
  \SC\gamma_a \SC^{-1} &=& \gamma_a{}^{\rm T}\,,\qquad \SC^{\rm T} =
  -\SC\nonumber\,, \qquad 
  \SC^\dagger = \SC^{-1} \ .
\end{eqnarray}  
For an $\SU(2)'$ doublet $\lambda^i$, the symplectic Majorana condition is defined by 
\be
(\overline\lambda_i)^{\rm T} = \SC\; \varepsilon_{ij}\lambda^j  \;.
\ee
Here, $\varepsilon_{ij}$ is an antisymmetric two-by-two matrix of the form
\be
(\varepsilon_{ij}) =
 \left( \begin{array}{cc} 0 & 1 \\ -1 & 0 \end{array} \right)  \;.
\ee
The indices $i,j, \ldots=1,2$ are raised and lowered according 
to the convention:
$\lambda_i := \lambda^j\varepsilon_{ji}$ and 
$\lambda^i = \varepsilon^{ij} \lambda_j$, where $\varepsilon^{ij} =
\varepsilon_{ij}$ and therefore 
$\varepsilon^{ik} \varepsilon_{kj} = - \delta^i_j$.

For a spinor $\zeta^\a, \ (\a=1,\cdots, 2r)$ in the fundamental representation of the $\USp(2r)$ $R$-symmetry group,
the  symplectic-Majorana condition reads as
\begin{equation}
  \label{eq:Majorana}
  \SC^{-1} \,\bar\zeta_\a {}^{\rm T}= \Omega_{\a\b}\,\zeta^\b\,,
\end{equation}
where $\Omega$ is the symplectic ${\rm USp}(2N)$ invariant tensor. For 
a more general spinor $\chi^{\a\b\cdots}{}_{\rho\sigma\cdots}$, the symplectic Majorana
constraint would read
\begin{equation}
  \label{eq:chi-Majorana}
  \SC^{-1} \,(\bar\chi_{\a\b\cdots}{}^{\rho\sigma\cdots})^{\rm T}= \Omega_{\a\g}\,
  \Omega_{\b\delta}\;\Omega^{\rho\xi}\,\Omega^{\sigma\zeta}\cdots
  \,\chi^{\g\delta\cdots}{}_{\xi\zeta\cdots} \, .
\end{equation}

\section{Details of the Killing spinor analysis \label{KillingDetails}}
In this appendix, we fill in some of the details of the Killing spinor analysis in the main text \S\ref{localizingsolns}.

One can take an orthonormal frame for the metric \eqref{metricansatz} as
\begin{eqnarray}
\begin{array}{l}
e^{1}=f(r) \; dt \, , \\
e^{2}=\frac{u(r)}{r}dt+r\; d\phi \, ,  \\
e^{3}=h(r)dr \, , \cr
e^{4}=\frac \ell2 d\theta \, , \cr
e^{5}=\frac  \ell2 \sin\theta\left[d\psi+\wt  A_t(r)dt+\wt  A_\phi(r)d\phi\right] \, .
\end{array}
\end{eqnarray}
With respect to this orthonormal frame, the spin connections are:
\begin{eqnarray}
\omega^{1}_{~2}&=&{1\over 4rf}\left({2u\over r} -u'\right)  dr \, , \cr
\omega^{1}_{~3}&=&\left[{1\over 2h}\left( {2u^2\over r^3  f}-{uu'\over r^2 f}-2f'  \right)+ 
(u\wt A'_\phi-r^2\wt A'_t)\frac{\ell^2\wt A_t \sin^2\theta}{8r^2fh}\right]dt  \, ,  \cr
 && \; +\left[ {1\over 2fh}\left({2u\over r} -u'\right)+(u\wt A'_\phi-r^2\wt A'_t) \frac{\ell^2 \wt A_\phi 
 \sin^2\theta}{8r^2fh}\right]d\phi+(u\wt A'_\phi-r^2\wt A'_t)\frac{ \ell^2\sin^2\theta}{8r^2fh}d\psi   \, ,  \cr
\omega^{2}_{~3}&=& -\left[\frac{u'}{2rh}+\frac{\ell^2 \wt A_t \wt A'_\phi\sin^2\theta }{8rh}\right]dt 
-\left[ \frac{1}{h} +\frac{\ell^2 \wt A_\phi \wt A'_\phi\sin^2\theta }{8rh}\right]d\phi
-\frac{\ell^2 \wt A'_\phi(r)\sin^2\theta}{8rh}d\psi  \, ,  \cr
\omega^{1}_{~5}&=&(u\wt A'_\phi-r^2\wt A'_t)\frac{\ell \sin\theta}{4r^2f}dr  \, , \cr
\omega^{2}_{~5}&=&-\frac{\ell \wt A'_\phi \sin\theta }{4r}dr  \, , \cr
\omega^{3}_{~5}&=&\frac{\ell\sin\theta}{4h}  (\wt A'_t dt+\wt A'_\phi d\phi)  \, , \cr
\omega^{4}_{~5}&=& \cos\theta\left[d\psi+\wt A_tdt+\wt A_\phi d\phi\right] \, .
\label{spinconnection}
\end{eqnarray}

The gravitino variation equations \eqref{keq} are:
\be \label{KillSpin2}
(\p_\mu-\tfrac14\omega_\mu^{ab}\gamma_{ab}+\Xi_\mu)\e=0\ ,
\ee
where
\be
\Xi_\mu=\tfrac{\i}{4} \,
  T_{ab}( 3\,\gamma^{ab}\gamma_\mu-\gamma_\mu\gamma^{ab}) =\tfrac{\i}2 T^{ab}e^c_\mu( \gamma_{abc}-4\delta_{ac}\gamma_{b})~,
\ee
the spin connections $\omega_\mu^{ab}$  \eqref{spinconnection} contain the derivative terms 
$\wt A'_\mu$ of the KK gauge fields.
To make use of the solutions in \cite{Izquierdo:1994jz}, we will need to cancel the 
derivative terms, which we shall do by changing the auxiliary two-form by $\delta T$ away 
from the ambient value \eqref{Toldvalue}.

Since an $S^2$ is fibered by rotating the $\psi$-direction, there is no change in the $\theta$-direction in the metric \eqref{metricansatz}. Imposing the projection condition \eqref{projcon1}, we have the following constraints on the auxiliary two-form $T_{ab}$ from $\Xi_\theta \e=0$. 
 \bea
\begin{array}{l}
\delta T^{15}= \delta T^{25}= \delta T^{35}=\delta T^{45}=0\ ,\cr
\delta T^{32}+2 \delta T^{14}=0\ ,\cr
\delta T^{13}+2 \delta T^{24}=0\ ,\cr
\delta T^{21}+2 \delta T^{34}=0\ .
\end{array}
 \eea
With these identities, one can write
\be
\Xi_\psi=\tfrac {3\i\ell}{4}\sin\theta\left[\delta T^{12}\gamma_{12} + 
\delta T^{13}\gamma_{13}+ \delta T^{23}\gamma_{23}\right]~.
\ee
One can choose the auxiliary two-form such that it cancels the $\wt A'$ pieces of the spin 
connections $\omega_\psi^{13}$ and $\omega_\psi^{23}$: 
\bea
\begin{array}{l}
\delta T^{12} = \delta T^{34}=0\ ,\cr
\delta T^{31} = 2 \delta T^{24}=\i(u\wt A'_\phi-r^2\wt A'_t)\frac{ \ell\sin\theta}{12r^2fh}\ ,\cr
\delta T^{23} = 2 \delta T^{14}=\frac{\i\ell \wt A'_\phi\sin\theta}{12rh}~.
\label{Taux}
\end{array}
\eea
With this choice of the auxiliary two-form, one can check, that no term proportional to $\wt A'$ survives 
in the gravitino variation \eqref{KillSpin2}.

The full value of the auxiliary two-form $T$ is obtained by collecting the 
configurations \eqref{Toldvalue} and \eqref{deviation}:
\bea\label{twoformaux}
T&=&\tfrac{\ell}{4}\sin\theta d\theta\wedge \left[d\psi+\wt A_tdt+\wt A_\phi d\phi\right]-\tfrac{\i\ell}{6}\sin\theta \left[ \wt A'_tdr\wedge dt+\wt A'_\phi dr\wedge d\phi\right]\cr
&&+\frac{\i\ell^2 \wt A'_\phi\sin\theta}{24 r f h}\left(1+\frac r\ell\sqrt{f^2-1}\right)dt\wedge d\theta -\frac{\ell^2\wt A'_\phi \sin\theta}{24  f h} \sqrt{f^2-1}d\phi\wedge d\theta \, . 
\eea

\section{Off-shell BPS solutions in $\AdS_{3}$ \label{ads3offshellsolns}}

In this appendix, we will solve the Killing spinor equations \eqref{killing}
\be
\left[\partial_\mu+\i B_\mu^a\gamma_a(\sigma_3\otimes\mathbbm1) -\tfrac {\i }2\wt A_\mu (\sigma_3\otimes\mathbbm1)\right]\zeta=0 \ , 
\label{killing2}
\ee
by following the methods in \cite{Izquierdo:1994jz}.
We only need a minor change of their strategy to construct solutions for the equations \eqref{killing2}. The difference comes from the fact that we work in the Euclidean signature instead of the Minkowski signature and we have
the extra factor $\sigma_3\otimes \mathbbm1$ in the equations due to the 5d lift. 

The equations \eqref{killing2} have the integrability condition
\be
\left[ S^\mu{}_a\gamma^a(\sigma_3\otimes\mathbbm1)-G^\mu (\sigma_3\otimes\mathbbm1)\right]\zeta =0 \ ,
\label{integrability}
\ee
where
\bea
S^\mu{}_a &\equiv& \varepsilon^{\mu\nu\rho}(\partial_\nu B_{\rho\, a} -
\varepsilon_{abc}B_\nu{}^b B_\rho{}^c)\ ,\cr
G^\mu &\equiv& \frac12\varepsilon^{\mu\nu\rho}\partial_\nu \wt A_\rho \ .
\eea
With the metric ansatz \eqref{metricansatz}, one finds that 
\be
S^r{}_1 = S^r{}_2 = S^t{}_3 = S^\phi{}_3 =0 \ , \ \ \ \ G^r=0 .
\ee
As in \cite{Izquierdo:1994jz}, we impose a projection condition
\be
(1-\Gamma^a b_a)\zeta=0\ ,
\label{projcon2dummy}
\ee
for some complex functions $b_a(r)$  satisfying
\be
b\cdot b=1 \ .
\ee
Here we take the 3d gamma matrices $\Gamma^a \ (a=1,2,3)$ as
\be
\Gamma^1=\sigma^3\otimes\sigma^1 \, , \qquad 
\Gamma^2=\sigma^3\otimes\sigma^2 \, , \qquad 
\Gamma^3=\mathbbm1\otimes\sigma^3 \; .
\label{3dgamma}
\ee 
The projection condition \eqref{projcon2dummy} on the Killing spinor $\zeta$ can be solved by writing
\be
\zeta=N (1+\Gamma^a b_a)\zeta_{{}_0}\ ,
\label{kspi}
\ee
for an arbitrary complex function $N(t,\phi,r)$ and a constant spinor $\zeta_{{}_0}$.
When \eqref{kspi} is substituted into the integrabilty condition
\eqref{integrability}, one finds that
\bea
S^\mu{}_3  b^3+(S^\mu{}_1  b^1+S^\mu{}_2  b^2- G^\mu) (\sigma_3\otimes\mathbbm1)+\left[S^\mu{}_3\gamma^3(\sigma_3\otimes\mathbbm1) + (\i \varepsilon_{3bc}S^{\mu\, b}b^c - G^\mu b_3)\gamma^3 \right  ]\cr
 +(S^\mu{}_1+\i \varepsilon_{1bc}S^{\mu\, b}b^c - G^\mu b_1)\gamma^1(\sigma_3\otimes\mathbbm1)+(S^\mu{}_2+\i \varepsilon_{2bc}S^{\mu\, b}b^c - G^\mu b_2)\gamma^2(\sigma_3\otimes\mathbbm1)  =0\ , \cr
\eea
which amounts to 
\be
S^\mu{}_3 = 0 \, , \qquad S^\mu{}_a+\i \varepsilon_{abc}S^{\mu\, b}b^c - G^\mu b_a = 0~.
\label{idS}
\ee
This leads to
\bea
&&B_{\phi2}B_{t1} = B_{\phi1}B_{t2} \; , \cr
&& (1-(b_1)^2) S^{t1} +(\i b_3 - b_2 b_1) S^{t2} =0 \; , \cr
&&(1-(b_1)^2) S^{\phi1} +(\i b_3 - b_2 b_1)S^{\phi2} =0\; .
\label{id}
\eea

We now return to the Killing spinor equations \eqref{killing2}. Given that the Killing spinor $\zeta$ takes
the form  of \eqref{kspi}, we find by substitution into \eqref{killing2} that
\bea
&&\partial_{t,\phi}(\ln N) +\i b_aB_{t,\phi}^{~a}(\sigma_3\otimes\mathbbm1)  -\frac {\i }2\wt A_{t,\phi}(\sigma_3\otimes\mathbbm1)   =0 \; , \cr
&&\partial_{r}(\ln N) +\i b_aB_{r}^{~a} =0 \; , 
\label{eqtphi}
\eea
and
\be
B_\mu{}^a - \i \varepsilon^{abc} b_b B_{\mu\, c} - \i \partial_\mu b^a
- b^ab_c B_\mu^c =0\ .
\label{second}
\ee\
The $r$ component of \eqref{second} yields
\bea
\i b_1' &=& -B_{r}^{~3}(\i b_2 + b_1 b_3) \; , \cr
\i b_2' &=& B_{r}^{~3}(\i b_1 - b_2 b_3) \; , \cr
\i b_3' &=& B_{r}^{~3}(1 - (b_3)^2) \; .
\label{r}
\eea
The $t$ and $\phi$ components of \eqref{second}   reduce to the algebraic equations
\bea
B_{\phi}^{~1} (1-(b_1)^2) +B_{\phi}^{~2}(-b_1 b_2 +\i b_3) &=&0 \; , \cr
B_{\phi}^{~1}  (-\i b_3 -b_1 b_2) +B_{\phi}^{~2}(1-(b_2)^2) &=&0 \; , \cr
B_{\phi}^{~1}  (\i b_2 - b_3 b_1) +B_{\phi}^{~2}(-b_3 b_2 -\i b_1) &=&0 \ .
\eea
Solving the second of these equations for $b_3$ and substituting it
into the other two equations, we find that
\be
b_1 = -Rb_2 \pm \sqrt{1+R^2} \, ,  \qquad b_3 = -\i [ R \mp b_2 \sqrt{1+R^2}]\ ,
\ee
where
\be
R = {B_{\phi }^{~2}\over B_{\phi}^{~1}}
=- {1\over 2f}\big( u' -{2u\over r}\big) + \frac{\i hr}{\ell} \ .
\label{eqR}
\ee
Substituting this result into the equations \eqref{r}, one discovers
that $R$  and
$b_2$ satisfy the ordinary differential equations
\be
R' - 2(1+R^2) B_{r}^{~3} =0 \; , 
\label{eqRpr}
\ee
and
\be
b_2 ' = \pm B_{r}^{~3}\sqrt{1+R^2} (1-b_2^2)\ .
\label{b2}
\ee

The equation for $R$ together with the first equation in  \eqref{id} is equivalent to the following two
equations that determine the functions $f$ and $h$ in terms of $u$ and
two  constants $\alpha$
and $\beta$:
\be 
\frac{2hf}{\ell} =-\i \left( 2\alpha + {u'\over r}\right) \, , \qquad 
f^2 = -\left(\alpha r + {u\over r}\right)^2 + \beta^2 \ .
\label{c20}
\ee
where we assume that $u(r)$ and $\alpha$ are purely imaginary. The constant $\alpha$ must be purely imaginary and cannot vanish if the spacetime is to be
asymptotic to anti-de Sitter space (or the black hole vacuum).

One can now deduce the following useful formulae:
\bea
R = {\i \sqrt{f^2-\beta^2} \over f}\ , \qquad
\sqrt{1+R^2} = {\beta\over f}\ , \qquad
B_{r}^{~3} = {1\over 2f} \big(\a r +{u\over r}\big)'  \ .
\eea
Using these formulae one can easily solve \eqref{b2} for $b_2$ and
simplify  the
expressions for the other components of $b_a$. The result is
\bea
b_1 &=&-{\i \sqrt{f^2-\beta^2}\over f}b_2 + {\beta\over f}\ ,\cr
b_2 &=&\i {k_+ \sqrt{f+\beta} + k_-\sqrt{f-\beta}\over
k_- \sqrt{f+\beta} - k_+\sqrt{f-\beta}}\ , \cr
b_3 &=&{\sqrt{f^2-\beta^2}\over f} + {\i \beta\over f} b_2 \ ,
\label{b}
\eea
where $k_\pm$ are arbitrary constants.

We can bring the function $N$ to the form
\bea
N = a(r,\phi) e^{({\i n\over 2}\phi +\frac {\i m}2\frac t\ell)(\sigma_3\otimes\mathbbm1)}\ ,
\eea
where $m, n$ are integers and $a(r,\phi)$ is a real function. Taking real and
imaginary  parts of \eqref{eqtphi}
now determines the one-form $\wt A$ and the function $a(r,\phi)$. It is
at this  point that one
discovers that $\beta$ must be purely real rather than purely
imaginary  because $a(r,\phi)$ would otherwise be a non-periodic
function  of $\phi$. For real $\beta$ one finds that
\be\label{gaugefieldsolnapp}
\wt A \= \wt A_tdt+\wt A_\phi d\phi 
\=  \Big({\i \beta u'\over \ell(2\a r +u')}+\frac{\i \beta}{\ell}+\frac m\ell\Big)dt +
\Big({2\i \beta r\over\ell (2\a r + u')} +{n}\Big)d\phi \; , 
\ee
and
\bea
a = a(r) \equiv k_+ \sqrt{f-\beta}- k_- \sqrt{f+\beta} \ .
\label{a}
\eea

Combining \eqref{kspi} with \eqref{b} now yields the following
result for the Killing spinor $\zeta$:
\bea
\zeta &=&e^{({\i n\over 2}\phi +\frac {\i m}2\frac t\ell)(\sigma_3\otimes\mathbbm1)}[k_+ \sqrt{f-\beta}- k_- \sqrt{f+\beta} ] \cr
& \times&\Bigg\{\Big[ 1 + {1\over f} \big( \beta\Gamma^1 +\sqrt{f^2-\beta^2}\Gamma^3\big)\Big] +
b_2\Gamma^2 \Big[ 1 - {1\over f} \big( \beta\Gamma^1 +\sqrt{f^2-\beta^2}\Gamma^3\big)\Big]\Bigg\}\zeta_0\ . 
\label{townsend solution2}
\eea
It is convenient to normalize the constant spinor $\zeta_0$
such that
\be \zeta_0^\dagger \zeta_0 =1\ .
\ee
In addition, we may require without loss of generality that $\zeta_0$
satisfy  $P\zeta_0=\zeta_0$
for some constant projection matrix, $P$, that projects out two
components of  $\zeta_0$, since
(for fixed $k_\pm$) the real dimension of the space of Killing spinors
$\e$ is two, not four,
despite the fact that $\zeta_0$ has two complex, and hence four real,
\be
\left[1+\tfrac{ (k_+^2 -k_-^2)\Gamma^1 + 2k_+k_-\,
\Gamma^3}{k_+^2 + k_-^2}\right]\zeta_0 = 0\ .
\ee

The metric admitting the Killing spinors \eqref{townsend solution2} is give by
\bea
ds^2 &=& \left[-(\a r)^2 -2\a u + \beta^2\right]dt^2 +2u dtd\phi
+\frac{\ell^2\left( \alpha + {u'\over 2r}\right)^2 }{\left(\alpha r + {u\over r}\right)^2 - \beta^2 }dr^2
+r^2d\phi^2 \cr
&+&\frac{\ell^2}{4}\left[d\theta^2+\sin^2\theta\left\{d\psi+ \Big({\i \beta u'\over \ell(2\a r +u')}+\frac{\i \beta}{\ell}+\frac m\ell\Big)dt +x
\Big({2\i \beta r\over\ell (2\a r + u')} +{n}\Big)d\phi\right\}^2\right]~.\cr
&&\label{5dmetric}
\eea
Notice that the requirement that there exist Killing spinors leaves
undetermined the purely imaginary function $u(r)$ and the constants $\a, \beta$. 
\\

We shall now summarize some useful identities which are relevant to the content of \S\ref{localizingsolns}. After changing $u\to \i u$ as done in \eqref{5d metric} and \eqref{gaugefieldsoln}, we find $-\i\ell \wt A_t+\wt A_\phi={\rm const}$ which implies $ \wt A'_\phi=\i\ell \wt A'_t $. The identity $ \wt A'_\phi=\i\ell \wt A'_t $ is used for the evaluation of the 3d CS term in \eqref{CS cubic}. In addition, by using \eqref{c20}, the field configuration in the middle of \eqref{deviation} can be rewritten as
\be
 \frac{ r^2\wt A'_t-\i u\wt A'_\phi}{r^2fh}=-\frac{\i \wt A'_\phi\sqrt{-1+f^2}}{rhf} \ .
\ee
From this result, one can show  
\bea 
\wt A'_t dr\wedge dt + \wt A'_\phi dr\wedge d\phi&=&-\frac{ r^2\wt A'_t-\i u\wt A'_\phi}{r^2fh}e^1\wedge e^3-\frac{ \wt A'_\phi}{rh}e^2\wedge e^3\cr&=&\frac{\wt A'_\phi}{rh}\left[\i \frac{\sqrt{-1+f^2}}{f}e^1\wedge e^3-e^2\wedge e^3\right]~,
\eea
which is relevant to \eqref{fs} and \eqref{twoformaux}.

\section{Evaluation of the action on the localizing solutions \label{actionapp}}
In this appendix, we shall explicitly evaluate the action \eqref{effective} on the BPS configurations found in \S\ref{localizingsolns}.

The Ricci scalar  of  the metric \eqref{5d metric} is 
\be\label{Ricciscvalue}
\CR=-\frac{2}{\ell^2}+\frac{\ell^2 (\wt A'_\phi)^2\sin^2\theta}{2(2r+\ell u')^2} + \frac{8\ell (u'-ru'')}{(2r+\ell u')^3}~.
\ee
Note that we use the convention of Wess and Bagger for curvature tensors where the Ricci scalar 
$\CR$ of a $\AdS$ space is positive (see appendix \S\ref{conventions} for details). 
The gauge field and auxiliary fields combine to give
\begin{small}
\be
3\,C_{IJK} \sigma^I\Big[\tfrac14 F_{\mu\nu}{}^J F^{\mu\nu K} - Y_{ij}{}^J  Y^{ijK }
  -3\,\sigma^J F_{\mu\nu}{}^K  T^{\mu\nu} \Big]+C(\sigma) 18T^2=C(\sigma)\left[- \frac{3}{\ell^2}+\frac{\ell^2 (\wt A'_\phi)^2\sin^2\theta}{4(2r+\ell u')^2}\right] \, .
\ee
\end{small}
The 5d Chern-Simons term $WFF$ in \eqref{effective} generates a 3d Chern-Simons term
\bea
 - \tfrac{\i}{64\pi^2}C_{IJK}\int W^I \wedge  F^J
\wedge F^K &= &\tfrac{\i \ell^3C(\sigma)}{16\pi}  \int \wt A d\wt A=\tfrac{\i \wt k}{4\pi}  \int \wt A d\wt A\cr
& =&\tfrac{\i \ell^3C(\sigma)}{8\pi}  \int \wt A'_\phi dt\wedge d\phi\wedge dr  \, .
\label{CS cubic}
\eea

Using these results, one can evaluate the bulk action integral 
\bea\label{bulkint}
{\cal S}_{\rm bulk}&=&\tfrac1{8\pi^2 } \int d^5x \sqrt{g}\left[ 3\,C_{IJK} \sigma^I\Big(\tfrac14 F_{\mu\nu}{}^J F^{\mu\nu K} - Y_{ij}{}^J  Y^{ijK } -3\,\sigma^J F_{\mu\nu}{}^K  T^{\mu\nu} \Big)\right. \cr
&& \hspace{4cm}\left.
  - C(\sigma) \Big(\tfrac12 \mathcal{R} - 18 T^2\Big)\right] - \tfrac{\i}{64\pi^2}C_{IJK}\int W^I \wedge  F^J
\wedge F^K \cr
&=&\tfrac{C(\sigma)}{8\pi^2 }\int  d^5x\tfrac{\ell^2}8(2r+\ell u')\sin\theta \left[ -\tfrac12 \tfrac4{\ell^2}  \right] \cr
&=&-\tfrac{C(\sigma)}{8\pi } \left[ r^2+\ell u(r)\right]^{\Lambda}_0 {\rm Area}(T^2_{\rm bdry}) \cr
&=&-\tfrac{C(\sigma)}{8\pi }\left[ \Lambda^2+\ell^2 u_\infty\right] {\rm Area}(T^2_{\rm bdry}) \; . 
\eea
where $T^2_{\rm bdry}$ denotes the boundary torus with coordinates \eqref{wtorus} 
on which the $(0,4)$ $\SCFT_2$ lives.
In the second line, we use the identities $\sqrt{g}=\tfrac{\ell^2}8(2r+\ell u')\sin\theta$ and 
$ \frac{u'-ru''}{(2r+\ell u')^2} = \frac{\wt A'_\phi}{2\ell}$. 
Since the action integral \eqref{bulkint} will diverge due to the infinite volume of $\AdS_3$, 
we regulate it by imposing a cutoff $r=\Lambda$. 
In the last line, we substitute the limits at infinity and the value at the origin $r=0$ of the function $u(r)$
\bea
 u(r) \stackrel{r \to \infty}{\longrightarrow}  \ell u_\infty\ , \quad  \ \ \ \ u(r) \stackrel{r \to 0}{\longrightarrow} 0 \ ,
\eea
where $u_\infty$ is a constant. These limits are compatible with the discussion in \S\ref{holomorphy}.

The boundary action has three pieces, which were discussed in \S\ref{ads3partfn}. 
Firstly, we have the Gibbons-Hawking term which ensures that, upon variation 
with fixed metric at the boundary, the action yields the Einstein equations
\bea
{\cal S}_{\rm GH}&=&\tfrac{C(\sigma)}{8\pi^2 }\int_{r=\Lambda} d^4x \sqrt{h}~K\cr
&=&\tfrac{C(\sigma)}{8\pi^2 }\int d^4x \sin\theta\left[ \tfrac{\Lambda^2}2 + \tfrac{\ell^2}4 (1+2u_\infty)
+{\cal O}(\tfrac1{\Lambda^2})\right]\cr
&=&\tfrac{C(\sigma)}{8\pi }  {\rm Area}(T^2_{\rm bdry})\left[2\Lambda^2 +\ell^2(1+2 u_\infty)\right]
+{\cal O}(\tfrac1{\Lambda^2}) \, , 
\eea
where $h_{\mu\nu}$ is the induced metric on the boundary and $K$ is the trace of the extrinsic curvature on the boundary, which are given by
\bea
\sqrt{h}&=&\tfrac{\ell}4\sin\theta\left[ r^2\ell^2+(r^2+\ell u(r))^2\right]^{1/2}\Big|_{r=\Lambda}=\tfrac\ell4 \sin\theta\left[ \Lambda^2\ell^2+(\Lambda^2+\ell^2 u_\infty)^2\right]^{1/2} \, , \cr
K&=&\left. \frac{2 r (\ell^2 + 2 r^2 + 2 \ell u) + 
 2 \ell (r^2 + \ell u) u'}{\ell (2 r + \ell u')\sqrt{
 r^2\ell^2+(r^2+\ell u)^2} }\right|_{r=\Lambda}\cr
&=&\left[ \Lambda^2\ell^2+(\Lambda^2+\ell^2 u_\infty)^2\right]^{-1/2}\tfrac1\ell\left[2 \Lambda^2+\ell^2(1+2u_\infty)\right]+{\cal O}(\tfrac1{\Lambda^3}) \, .
\eea
To produce a finite result, one also needs to add the boundary cosmological constant 
(this term and the Gibbons-Hawking term were already indicated in \eqref{Sgrav}): 
\bea
{\cal S}_{\rm ct}&=&-\tfrac{C(\sigma)}{8\pi^2 }\int_{r=\Lambda} d^4x \sqrt{h}~\tfrac1\ell\cr
&=&-\tfrac{C(\sigma)}{8\pi^2 }\int d^4x\tfrac14\sin\theta\left[ \Lambda^2\ell^2+(\Lambda^2+\ell^2 u_\infty)^2\right]^{1/2} \cr
&=&-\tfrac{C(\sigma)}{8\pi^2 }\int d^4x\sin\theta\left[ \tfrac{\Lambda^2}4 + \tfrac{\ell^2}8 (1+2u_\infty)+{\cal O}(\tfrac1{\Lambda^2})\right]\cr
&=&-\tfrac{C(\sigma)}{8\pi }  {\rm Area}(T^2)\left[\Lambda^2 +\tfrac{\ell^2}2(1+2 u_\infty)\right]+{\cal O}(\tfrac1{\Lambda^2})~.
\eea
So far, the evaluation of the action reduces to 
\be
{\cal S}_{\rm bulk}+{\cal S}_{\rm GH}+{\cal S}_{\rm ct}
=\tfrac{\ell^2C(\sigma)}{16\pi } {\rm Area}(T^2_{\rm bdry}) \, .
\ee
Substituting the various constants \eqref{relconsts}, 
we get:
\be
{\CS}_{\rm bulk}+{\cal S}_{\rm GH}+{\cal S}_{\rm ct}
= -\frac{i\pi c}{12 \ell}(\tau-\bar\tau) \; .
\ee

\bibliography{5DSUGRA}{}
\bibliographystyle{JHEP}

\end{document}